\documentclass[10pt,journal,twoside,final]{IEEEtran}

\usepackage[switch]{lineno} 
\usepackage{blindtext}
\usepackage{graphicx}
\usepackage{cite}
\usepackage{subfigure}
\usepackage{amssymb,amsmath}
\usepackage{amsthm}
\usepackage{mathtools}
\usepackage{multirow}
\usepackage{multicol}
\usepackage{verbatim}
\usepackage{color, soul}
\usepackage{enumitem} 
\usepackage[normalem]{ulem}
\usepackage{lipsum}
\usepackage{cuted}
\usepackage{bm}
\usepackage{Maths}
\usepackage{adjustbox}    
\usepackage{threeparttable}
\usepackage{booktabs}
\usepackage{array,tabularx,adjustbox,threeparttable, xparse}
\usepackage[table]{xcolor}
\usepackage[colorlinks=true,linkcolor=black,urlcolor=black,citecolor=blue]{hyperref}
\newcolumntype{R}[1]{>{\raggedright\arraybackslash}p{#1}}
\usepackage[most]{tcolorbox}
\usepackage{tikz}

\usepackage{dirtytalk}
\usepackage[normalem]{ulem}

\usepackage{algorithm}
\usepackage{algorithmic}

\newcommand{\stagebreak}{\noindent\tikz{\draw[gray!60,line width=0.4pt] (0,0) -- (\linewidth,0);} \vspace{1pt}}
\begin{document}
	\setlength\linenumbersep{10pt} 
	\modulolinenumbers[1]         
	\title{Near-Optimal Reconfigurable Intelligent Surface Configuration: Blind Beamforming with Sensing}
	\author{
		Son~Dinh-Van, \IEEEmembership{Member,~IEEE},
		Nam~Phuong~Tran, 
		and Matthew~D. Higgins, \IEEEmembership{Senior Member,~IEEE}
	\IEEEcompsocitemizethanks{
		\IEEEcompsocthanksitem
		This work was supported in part by the WMG Centre High Value Manufacturing Catapult (HVMC), University of Warwick, Coventry, U.K., and in part by the UK ISPF programme through the UKRI EPSRC (grant ID: UKRI554, led by University of East Anglia) under pilot project ‘BEAM-RAN’ (UEA ref: R213867). An early version of this paper was presented in part at the IEEE International Conference on Communications Workshops (ICC), Glasgow, U.K., May 2026; \textit{doi: 10.1109/ICCWorkshops63917.2026.11586738.} Son~Dinh-Van and Nam Phuong Tran contributed equally to this work. \textit{(Corresponding author: Son~Dinh-Van.)}
		\IEEEcompsocthanksitem
		Son~Dinh-Van and Matthew D. Higgins are with Warwick Manufacturing Group, University of Warwick, Coventry CV4 7AL, U.K. (e-mail: {son.v.dinh, m.higgins}@warwick.ac.uk).
		\IEEEcompsocthanksitem
		Nam Phuong Tran was with the Department of Computer Science, University of Warwick, Coventry CV4 7AL, U.K. He is now with Inria Centre at the University of Lille, 59650 Villeneuve-d'Ascq, France (e-mail: phuong-nam.tran@inria.fr).
	}
	}
	\maketitle
	\thispagestyle{empty}
	\pagestyle{empty} 
	\begin{abstract}
		Blind beamforming has emerged as a promising approach to configure reconfigurable intelligent surfaces (RISs) without relying on channel state information (CSI) or geometric models, making it directly compatible with commodity hardware. This paper considers a single-input single-output (SISO) communication system aided by a binary RIS. We propose a new blind beamforming algorithm, so-called Blind Optimal RIS Beamforming with Sensing (\textsc{BORN}), to find a phase configuration that maximizes the received signal-to-noise ratio (SNR) using received signal strength (RSS) measurements only. In contrast to existing methods that rely on majority-voting mechanisms, \textsc{BORN} exploits the intrinsic quadratic structure of the received SNR. The algorithm proceeds in two stages: \emph{sensing}, where a quadratic model is estimated from RSS measurements, and \emph{optimization}, where the RIS configuration is obtained using the estimated quadratic model. Our novelties are twofold. Firstly, we show for the first time that, despite using RSS measurements only, \textsc{BORN} achieves a near-optimal performance relative to the case where perfect channel state information is available, using only $O\big(N\log_2 N\big)$ samples, where $N$ is the number of RIS elements. As a by-product of our analysis, we show that quadratic models are learnable under Rademacher feature distributions when the second-order coefficient matrix is low-rank. This result, to our knowledge, has not been established in prior matrix sensing literature. Secondly, extensive simulations and real-world field tests demonstrate that \textsc{BORN} achieves near-optimal performance, substantially outperforming state-of-the-art blind beamforming algorithms, particularly in scenarios with a weak background channel such as non-line-of-sight (NLOS).
	\end{abstract}
	\section{Introduction}
	Reconfigurable Intelligent Surfaces (RISs) are emerging as a transformative technology for sixth-generation (6G) wireless networks \cite{9998527}. By utilizing a large array of low-cost, passive reflecting elements, RISs enable control over reflected signal paths, thereby enhancing signal strength, coverage, and energy efficiency in wireless communications. Due to their flexibility, cost-effectiveness, and energy-efficient design, RISs offer a compelling solution for addressing propagation challenges without requiring additional active infrastructure, such as extra access points. For instance, in manufacturing environments where heavy machinery often creates signal blockages, RISs can ensure connectivity for connected workers, facilitating seamless real-time monitoring and coordination in automated production lines~\cite{9761230}.
	
	Effective RIS configuration requires the system to acquire information about the wireless environment. Such information can be obtained in different forms, including full or partial channel state information (CSI), side information such as user location or channel statistics, or simple receiver measurements such as received signal strength (RSS). In this broad sense, RIS control relies on a form of sensing to acquire communication-related information from the environment. Note that the term “\textit{sensing}” here is different from radar sensing, where the objective is environmental or target sensing~\cite{Magbool2025Survey}. To control the RIS, a widely adopted approach is to sense the channel through explicit CSI estimation and then optimize the RIS phase shifts accordingly~\cite{9779545,10820854,10570724}. However, from a practical standpoint, this method is often difficult to realize in RIS-assisted systems due to weak reflected links, substantial protocol overhead, and the limited availability of phase measurements on commercial hardware~\cite{arun2020rfocus,9940989,10694808, yan2025power}. Other forms of sensing for RIS-assisted communications have also been considered, such as statistical channel information~\cite{9761230,gan2021risstatcsi} or accurate user locations~\cite{9779399,zheng2025locationdriven,10070578}, but these auxiliary inputs may be unavailable or unreliable in practice. This motivates an alternative strategy based solely on RSS measurements, which avoids explicit CSI estimation and reduces the dependence on auxiliary information. Blind beamforming has therefore emerged as an attractive direction, enabling RIS configuration directly from RSS measurement feedback without requiring explicit CSI, transceiver position information, or directional channel models.
	
	Most advanced blind beamforming techniques rely on statistical methods, wherein the system tries out a number of phase shift configurations, and the receiver reports corresponding RSS measurements~\cite{arun2020rfocus, 9940989, 10694808}. These data are then processed to find a good RIS configuration. To evaluate the effectiveness of a blind beamforming algorithm, two aspects are particularly important: \textit{(1) Sample complexity:} How many samples are required to achieve satisfactory performance? \textit{(2) Performance guarantees:} What guarantees can be established for the achieved performance? In this study, we consider a RIS-assisted wireless communication link in a single-input single-output (SISO) setting. We focus on the downlink, where a transmitter sends a signal to a receiver with the assistance of the RIS, and the design objective is to maximize the received signal-to-noise ratio (SNR) at the receiver using RSS measurements only. We propose a blind beamforming algorithm, so-called \textbf{\underline{B}}lind \textbf{\underline{O}}ptimal \textbf{\underline{R}}IS beamforming with se\textbf{\underline{N}}sing (\textsc{BORN}), that achieves near-optimal performance with high probability for binary RIS. Unlike existing approaches, which rely on majority-voting mechanisms and overlook the inherent structure of the received SNR, \textsc{BORN} exploits the fact that, in the binary setting, the received SNR represents a quadratic model (QM) with a Rademacher feature distribution. After data collection, \textsc{BORN} operates in two stages: \textit{(1) Sensing}, which estimates the parameters of QM representing the structure of received SNR; \textit{(2) Optimization,} which determines the phase shift configuration that maximizes the received SNR based on the QM learned in the sensing stage.
	\vspace{-0.2cm}
	\subsection{Contributions}
	The main contributions of this study are outlined as follows:
	\begin{itemize}
		\item We propose \textsc{BORN}, a blind beamforming algorithm, that configures RIS phase shifts using only RSS, thereby eliminating the need for explicit channel estimation. This makes \textsc{BORN} readily applicable to commodity hardware. Unlike most existing blind beamforming algorithms that rely on majority-voting mechanisms, \textsc{BORN} adopts a fundamentally different approach by leveraging the quadratic structure of the received SNR, enabling near-optimal performance.
		\item We also provide theoretical sample-complexity and performance guarantees, showing that \textsc{BORN} achieves an $O(\varepsilon)$ suboptimality gap with only $O\big(N\log_2(N/\varepsilon)\big)$ training samples, where $N$ is the number of RIS elements. To our knowledge, this is the first result that proves a suboptimality gap for binary RIS. In addition, we show that the end-to-end computational complexity scales quadratically with $N$.
		\item We present both experimental and simulation results. The field test was conducted in a manufacturing scenario. The results show that most existing blind beamforming algorithms degrade significantly when the background channel is weak. In contrast, with RSS measurements only, \textsc{BORN} consistently achieves near-optimal performance relative to the case when perfect CSI is available across all tested scenarios.
		\item As a by-product of our analysis, we prove that a QM is learnable under Rademacher feature distributions when its quadratic coefficient matrix satisfies the low-rankness and incoherence property. This overturns the prevailing belief that QMs are unlearnable under a Rademacher distribution \cite{lin2017second}. To the best of our knowledge, this is the first positive learnability result for this setting in the matrix sensing literature, and it is of independent interest.
	\end{itemize}
	\subsection{Related Works}
	A large body of literature considers RIS configuration under explicit or implicit knowledge of the propagation environment. In particular, many works assume perfect CSI and optimize the RIS coefficients accordingly, typically via alternating optimization, convex approximation, or machine learning~\cite{9779545,10820854,10570724}. Despite their effectiveness, such approaches rely on CSI acquisition, which is often challenging in practice~\cite{arun2020rfocus,9940989,10694808,yan2025power}. First, the reflected channels associated with individual RIS elements are often too weak to be reliably estimated in the presence of background noise. Second, CSI estimation for RIS usually requires substantial modifications to existing wireless protocols. In current 4G/5G systems, pilot signals are primarily designed for estimating the direct channels between base station and user, rather than the channels involving the RIS. Third, most commercial hardware only provide received signal strength indicators (RSSI), without the phase information needed for channel estimation. To reduce the burden of instantaneous CSI estimation, other works exploit side information such as statistical CSI~\cite{9761230,gan2021risstatcsi}. For example, \cite{9761230} utilizes unsupervised deep learning to configure the RIS based on large-scale channel parameters, without requiring labeled training data. While these approaches can be effective in their respective settings, they rely on prior knowledge, which is often not available.

	In blind beamforming, RIS configuration is performed using RSS measurements only. Compared with CSI acquisition, RSS is much easier to obtain, since it is readily available on most commercial wireless hardware through standard measurements such as the reference signal received power (RSRP). Notable blind beamforming algorithms include \textsc{RFocus}~\cite{arun2020rfocus}, Conditional Sample Mean (\textsc{CSM})~\cite{9940989}, Grouped Conditional Sample Mean (\textsc{GCSM})~\cite{10694808}, and Majority-Voting Conditional Sample Mean (\textsc{MV-CSM})~\cite{10612784}. One of the earliest studies in this area, \textsc{RFocus}, adjusts the state of each reflecting element, either $\texttt{ON}$ or $\texttt{OFF}$, using a majority-voting mechanism to maximize the RSS~\cite{arun2020rfocus}. Experimental results show that \textsc{RFocus} achieves approximately $50~\%$ of the maximum attainable RSS improvement with several thousand samples. In comparison, \textsc{CSM} utilizes conditional sample means to control the phase shifts. For non-binary RIS (e.g., $K > 2$), where $K$ is the number of discrete phase shifts per element, \textsc{CSM} guarantees a quadratic SNR boost~\cite{9940989}. Our extensive experiments show that both \textsc{RFocus} and \textsc{CSM} perform relatively well in line-of-sight (LOS) conditions, but degrade significantly in non-line-of-sight (NLOS) scenarios. This is also demonstrated in~\cite{10694808}. To address this limitation, \textsc{GCSM}, an incremental improvement over \textsc{CSM}, was proposed in \cite{10694808}. The core idea of \textsc{GCSM} is to partition the RIS elements into three groups, each randomly configured to emulate a virtual direct path, thus improving robustness in scenarios where the direct path is weak. Subsequently, the phase shifts for one group are determined at a time using \textsc{CSM}. Owing to the virtual direct path, \textsc{GCSM} is robust in both LOS and NLOS conditions. It guarantees at least a $\cos^2(\pi/K)$ fraction of the optimal performance, despite a higher sample complexity than \textsc{CSM}. In \cite{10612784}, \textsc{MV-CSM}, an extension of \textsc{CSM}, is proposed to improve the coverage enhancement for multiple users by allowing each user to vote for its preferred phase shift. Related to blind beamforming, \cite{yan2025power} focuses on channel autocorrelation matrix estimation using RSS for RIS-assisted communications. This research is closely related in spirit to ours, but it focuses on covariance estimation rather than the sensing-optimization framework considered in this paper.
	
	Although experimental and theoretical performance analyses have been established for non-binary RIS, prior studies provide no theoretical performance guarantees for binary RIS. This is because in the binary case, a clockwise phase rotation is identical to a counterclockwise rotation~\cite{9940989}. As a result, \textsc{CSM} cannot theoretically guarantee an SNR gain in the worst case~\cite{9940989}, and other algorithms relying on a similar mechanism also fail for this reason. For \textsc{GCSM}, $\cos^2(\pi/K) = 0$ when $K=2$, and thus no performance guarantee can be established. At the time of this work, binary RIS remains prevalent in commercial prototypes~\cite{Greenerwave2025,TMYTEK2025} for two main reasons~\cite{9779545}: (i) supporting more phase-shift levels requires substantially more electronic components, thereby increasing hardware cost; and (ii) implementing higher-order phase shifts necessitates additional PIN diodes, leading to increased reflection loss. Binary RIS is significantly cheaper and easier to design, at the cost of a few dB compared with higher-resolution quantized or ideal infinite-resolution phase shifts. In particular, ETSI reports that 1-bit and 2-bit phase quantization lead to approximately $3.9$~dB and $0.9$~dB received-power losses, respectively~\cite{ETSI_GR_RIS_004_2025}. Overall, despite the widespread use of binary RIS in practice, no existing blind beamforming method offers provable performance guarantees. Addressing this is vital for enabling robust operation in real-world deployments. 
	
	The QM is a fundamental regression class that augments a linear predictor with a quadratic term that captures interactions between features. It enjoys a wide range of applications including quantum tomography, recommendation systems, and click-through prediction. As an umbrella, QM covers phase retrieval \cite{candes2015phase}, symmetric rank-one matrix sensing \cite{kueng2017low, cai2015rop, zhong2015efficient}, and interaction models such as factorization machines \cite{rendle2010factorization}. As a result, the theoretical foundations for efficient learning QMs are still evolving. More recently, \cite{wang2025low} developed an analytical method for recovering a low-rank positive semi-definite (PSD) covariance matrix from structured rank-one measurements. However, it focuses on PSD covariance recovery under specially designed structured measurements and does not consider an additional linear term. When the feature distribution is Gaussian, properties such as rotational invariance and isotropy yield clean concentration and identifiability, so the problem is comparatively well understood. These properties enable strong guarantees for both convex lifting (nuclear-norm programs) and efficient nonconvex solvers \cite{kueng2017low, zhong2015efficient}. By contrast, for general sub-Gaussian distribution, learnability depends heavily on higher-order moments. The study proposed in \cite{lin2017second} shows that when the feature distribution satisfies a Moment Invertible Property (MIP), a mild relation between skewness and kurtosis, one can learn QM globally and at a linear rate via a moment-estimation-sequence (MES) procedure. Nevertheless, when the distribution is non-MIP, MES cannot eliminate the systematic gradient bias induced by third- and fourth-order moments. In this case, a diagonal-free structure becomes necessary and sufficient for consistent recovery \cite{lin2017second}. In this paper, our approach involves learning a QM efficiently when the feature vectors form a Rademacher distribution. Unfortunately, there is very limited research on this topic at the time of writing this paper. It is also important to mention that a Rademacher distribution does not satisfy MIP. In the absence of a diagonal-free property, it has been widely believed that QM is not learnable (see Proposition 1 of \cite{lin2017second}). We revisit this gap and show that QM can in fact be learned under Rademacher distribution when an appropriate low-rank structure is present.

	\textit{The rest of the paper is organized as follows.} Section~\ref{sec:system_model} introduces the system model and problem formulation. Section~\ref{sec:born} outlines \textsc{BORN}, with the subsequent sections detailing its components. Section~\ref{sec:stage1} develops the algorithm for the sensing stage, while Section~\ref{sec:optimization} presents the optimization stage. Section~\ref{sec:theoretical_analysis} establishes the theoretical guarantees of \textsc{BORN}. Section~\ref{sec:experiments} reports the performance benchmarks through simulations and field measurements. Finally, Section~\ref{sec:conclusion} concludes the paper.
	
	\textit{Notations in the main paper:} For $x\in\mathbb{C}$, $\Real{x}$ and $\Imaginary{x}$ represent the real and imaginary parts, respectively. Moreover, $\|\cdot\|_2$ is the Euclidean norm for vectors and the spectral norm for matrices, while $\|\cdot\|_F$ is the Frobenius norm. We use $\bm 1$ for the all-ones vector and $[N] \triangleq \{1,\dots,N\}$. For a matrix $\bm X$, $\bm X^\top$ and $\bm X^+$ are its transpose and Moore--Penrose pseudoinverse, respectively. For a symmetric matrix $\bm X$, $\mathrm{SVD}(\bm X,r)$ denotes the matrix whose columns are the orthonormal eigenvectors associated with the $r$ largest singular values of $\bm X$. In addition, $\mathrm{QR}(\bm X)$ denotes the Q-factor of the QR decomposition of $\bm X$, $\sigma_r(\bm X)$ is the $r$-th largest singular value of $\bm X$, and $\mathrm{sgn}(\cdot)$ is the sign function.	\textit{Notations specific to Appendix are defined therein.}
	\section{System Model and Problem Formulation}
	\label{sec:system_model}
	\begin{figure}[t]
		\centering
		\includegraphics[width=0.4\textwidth]{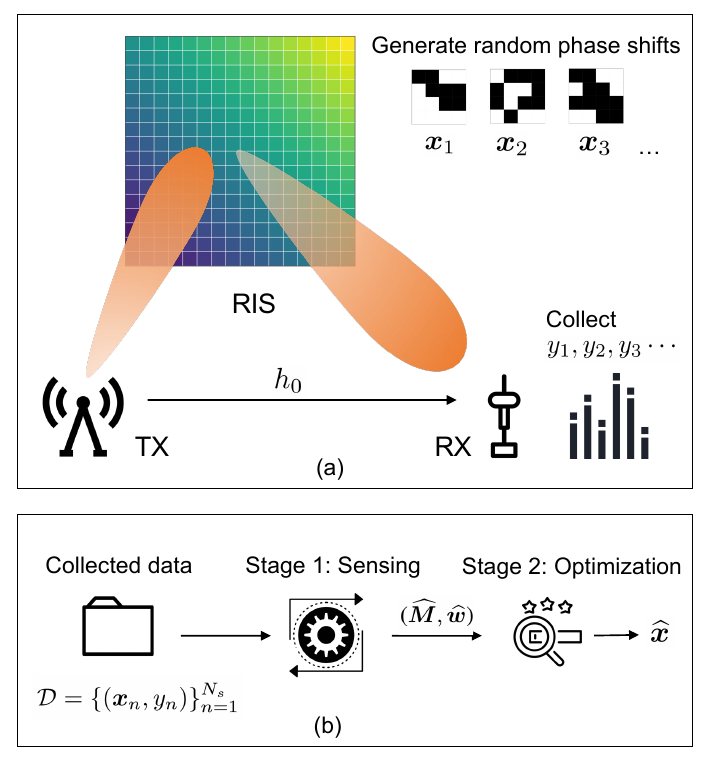}
		\caption{(a) The system model; and (b) the protocol of \textsc{BORN}.}
		\label{fig:system_model}
	\end{figure}
	In this study, we consider a wireless communication system comprising a transmitter (TX), a receiver (RX), and a reconfigurable intelligent surface (RIS) that reflects signals from the TX to the RX, as illustrated in Fig. \ref{fig:system_model}(a). The RIS consists of $ N $ passive reflective elements, each operating in two states.
	\vspace{-0.2cm}
	\subsection{Signal Model}
	Let $h_n$, for $n = 1, \dots, N$, denote the cascaded channel from the TX to RX induced by the $n$-th reflective element, and let $h_0$ represent the superposition of all those channels from the TX to RX that are not related to the RIS, namely the \textit{background channel}. The RIS phase setting is defined as $\bm{\theta} = \left[ \theta_1, \dots, \theta_N\right]^\top $, where $\theta_n$ denotes the phase shift induced by the $n$-th reflective element. For the binary RIS, $\theta_n \in \{ 0, \pi\}$.
	
	Let $X$ be the transmit signal with average power $P$, i.e., $\mathbb{E}[|X|^2] = P$. The received signal at the RX is given by\footnote{In practice, the hardware can exhibit amplitude-phase coupling. This phenomenon can be captured by using two complex coefficients per element, and the received SNR remains a quadratic function of the binary configuration; thus, the proposed framework remains applicable.}
	\begin{equation}
		Y = \left( h_0 + \sum_{n=1}^N h_n e^{j \theta_n} \right) X + Z,
	\end{equation}
	where $Z \sim \mathcal{CN}(0, \sigma^2)$ is a circularly symmetric complex Gaussian noise term modeling additive thermal noise. The received SNR\footnote{In most commodity transceivers, such as Wi-Fi, Bluetooth, and cellular chipsets, SNR is not directly reported. Instead, these devices provide the Received Signal Strength Indicator (RSSI), a standard metric output by the hardware \cite{8013036}. RSSI can be used to estimate SNR when the noise power is accurately measured.} can be computed as
	\begin{equation}
		s(\bm{\theta}) = \frac{P}{\sigma^2} \bigg| h_0 + \sum_{n=1}^N h_n e^{j \theta_n} \bigg|^2.
	\end{equation}
	When there is no RIS, the baseline SNR is $\mathtt{SNR}_0 = \frac{P |h_0|^2}{\sigma^2}.$
	\vspace{-0.4cm}
	\subsection{Problem Formulation}
	\label{sec:problem_formulation}
	In the context of blind beamforming, the channel coefficients $h_i$ are unknown, and only observations such as $s(\bm{\theta})$ and $\mathtt{SNR}_0$ are available. The goal is to identify an optimal phase configuration $\bm{\theta}^\star \in \{0, \pi\}^N$ that maximizes the SNR. To do this, the system evaluates $N_s$ independent and identically distributed (i.i.d.) probing phase configurations, $\bm{\theta}_1,\dots,\bm{\theta}_{N_s}$, where each entry $\theta_{i,n}$ is independently chosen from $\{0,\pi\}$ with equal probability, and measures the corresponding SNR values, $s(\bm{\theta}_1),\dots,s(\bm{\theta}_{N_s})$. These measurements are then used to determine the optimal configuration $\bm{\theta}^\star$, formulated as:
	\begin{equation}
		\begin{aligned}
			\mathrm{(P0)} \quad \underset{\bm{\theta}}{\mathrm{maximize}} \quad & s(\bm{\theta}) \\
			\mathrm{subject \ to} \quad & \bm{\theta} \in \{0, \pi\}^N.
		\end{aligned}
	\end{equation}
	Now, let $\bm{x} \triangleq [x_1, \ldots, x_N]^\top$ with $x_n = e^{j\theta_n}$ for $n = 1,\ldots,N$. Since $\theta_n \in \{0, \pi \}$, $x_n \in \{-1, 1\}$ and $\bm{x}$ is a Rademacher vector. Let $\bm{v}_n \triangleq [\Real{h_n}, \Imaginary{h_n}]^\top$ with $n=0,1,\cdots, N$. In addition, denote $\bm{V} \triangleq [\bm{v}_1,...,\bm{v}_N] \in \mathbb R^{2\times N}$.
	
	\begin{lemma}
		\label{lm: P1}
		In the real domain, the received SNR can be rewritten as \footnote{
		Equivalently, one may write
		\(s(\bm{x})=\tilde{\bm{x}}^\top\tilde{\bm{M}}_\star\tilde{\bm{x}}\),
		where \(\tilde{\bm{x}}=[1,\bm{x}^\top]^\top\),
		\(\tilde{\bm{M}}_\star=\frac{P}{\sigma^2}\tilde{\bm{V}}^\top\tilde{\bm{V}}\), and
		\(\tilde{\bm{V}}=[\bm{v}_0,\bm{V}]\).
		This form may be useful for alternative analyses and findings.
		}
		\begin{align}
			s(\bm{x}) = \bm{x}^\top \Mtrue \bm{x} + \bm{x}^\top \wtrue + \ctrue,
		\end{align}
		where $ \Mtrue \triangleq \frac{P}{\sigma^2} \bm{V}^\top \bm{V}$, $ \wtrue \triangleq 2 \frac{P}{\sigma^2} \bm{V}^\top \bm{v}_0$ and $\ctrue \triangleq \frac{P}{\sigma^2} \| \bm{v}_0\|_2^2$.
	\end{lemma}
	\begin{proof}
		Lemma \ref{lm: P1} is obtained via a simple transformation.
	\end{proof}
	According to Lemma 1, the received SNR can be expressed as a quadratic function of the Rademacher vector $\bm{x}$. Note that $r \triangleq \Rank \Mtrue \leq 2$. Moreover, $\ctrue = \mathtt{SNR}_0$, hence, it is measurable. Let us define $y(\bm x) \triangleq s(\bm{x}) - \ctrue$, then $y$ is also observable\footnote{In practice, RSS measurements may include external interference in addition to thermal noise. Since our method relies only on RSS, the RX can estimate the interference-plus-noise floor by muting the TX ($P=0$) and subtract it from the RSS readings to obtain the signal-only measurements used by our quadratic model; the remainder of the framework is unchanged.}. Therefore, $\mathrm{(P0)}$ is equivalent to the following optimization problem:
	\begin{equation}
		\begin{aligned} \label{eq: sys model in real domain}
			\mathrm{(P1)} \quad \underset{\bm{x}}{\mathrm{maximize}} \quad & y(\bm x) = \bm{x}^\top \Mtrue \bm{x} + \bm{x}^\top \wtrue, \\
			\mathrm{subject \ to} \quad & \bm{x} \in \{ -1, 1\}^N.
		\end{aligned}
	\end{equation}
	Now, our objective is to solve $\mathrm{(P1)}$, given $N_s$ instances $\{\bm{x}_1, \cdots, \bm{x}_{N_s}\}$ and the corresponding labels $\{y_1, \cdots, y_{N_s}\}$. This is challenging due to the unknown parameters $\Mtrue$ and $\wtrue$. In addition, $\bm{x}$ is constrained to the vertices of an $N$-dimensional hypercube, which makes solving $\mathrm{(P1)}$  computationally intractable due to the combinatorial nature of the feasible set, which classifies the problem as NP-hard.
	
	\begin{figure}[t]
		\centering
		\resizebox{0.9\linewidth}{!}{
			\begin{tcolorbox}[width=\linewidth,colback=gray!0,colframe=gray!100,
				boxrule=0.15pt,arc=1mm,left=0.8mm,right=0.8mm,top=0.6mm,bottom=0.6mm]
				\textbf{\textsc{BORN} Protocol}\hfill \textbf{Output:} $\bm{\theta}^\star$
				\stagebreak
				$\mathtt{Stage \ 0: \ Data \ Collection}$
				\begin{enumerate}[label=\textbf{0.\arabic*},leftmargin=*,itemsep=0.5pt,topsep=0.5pt]
					\item Measure $\mathtt{SNR}_0$.
					\item Draw $N_s$ i.i.d. probing configurations with $\theta_{i,n}$ chosen randomly from $\{0, \pi\}$, and record the corresponding received SNR. Next, construct a dataset $\mathcal{D}$, as described in Section \ref{sec:problem_formulation}.
				\end{enumerate}
				
				\stagebreak
				$\mathtt{Stage \ 1: \ Sensing}$
				\begin{enumerate}[label=\textbf{1.\arabic*},leftmargin=*,itemsep=0.5pt,topsep=1pt]
					\item Run Algorithm~\ref{alg: gFM} on $\mathcal{D}$ to obtain $(\Mest,\west)$.
				\end{enumerate}
				
				\stagebreak
				$\mathtt{Stage \ 2: \ Optimization}$
				\begin{enumerate}[label=\textbf{2.\arabic*},leftmargin=*,itemsep=1pt,topsep=2pt]
					\item Run Algorithm~\ref{alg:QuadraticOptimizing} using $(\Mest,\west)$ to obtain $\bm{\theta}^\star$.
				\end{enumerate}
			\end{tcolorbox}
		}
		\caption{A summary for the protocol of \textsc{BORN}.}
		\label{fig:protocol-born}
	\end{figure}
	\section{The Proposed Blind Beamforming Algorithm} \label{sec:born}
	In this section, we introduce \textsc{BORN}, an algorithm designed to solve $\mathrm{(P1)}$. Similar to other existing methods~\cite{arun2020rfocus,9940989,10694808,10612784}, \textsc{BORN} begins with a data collection stage, as discussed in Section \ref{sec:system_model}. This stage only requires the measurement of RSS, hence, it is compatible with commodity hardware. Regarding signal processing, the core idea is to learn the QM in a data-driven manner and then, we utilize this learned model to compute the optimal phase reconfiguration. To this end, the core signal processing of \textsc{BORN} consists of two important stages: \textit{sensing}, followed by \textit{optimization}.
	
	The sensing stage estimates the parameters $(\Mtrue, \wtrue)$ using the dataset $\mathcal{D} \triangleq \{(\bm{x}_{n}, y_{n})\}_{n=1}^{N_s}$. The algorithm utilized for this stage, so-called $\mathtt{MatrixSensing}$, is summarized in Algorithm~\ref{alg: gFM}. Once an estimate is obtained, \textsc{BORN} can proceed to the optimization stage, where $\mathrm{(P1)}$ is solved by substituting the estimated parameters in place of the unknown $(\Mtrue, \wtrue)$. The algorithm utilized for this stage, namely $\mathtt{QuadraticOptimizing}$, is outlined in Algorithm~\ref{alg:QuadraticOptimizing}. The rationale is that, if the estimation is sufficiently accurate, the underlying SNR structure is revealed, we can hope for a near-optimal phase configuration. The sensing stage is most accurate when the channel remains approximately stationary (or quasi-static) over the sensing window. Let $\tau_s$ denote the total sensing time. Two operating regimes can be distinguished: (i) \emph{adaptive control}, where $\tau_s$ should be shorter than the channel coherence time, and (ii) \emph{slow-timescale RIS control}, where the RIS is updated less frequently and the learned model can be interpreted as an effective model over the sensing window, provided that the channel statistics remain stationary over $\tau_s$.
	
	The complete protocol of \textsc{BORN} is illustrated in Fig.~\ref{fig:system_model}(b) and Fig.~\ref{fig:protocol-born}. The details of the two stages are elaborated in the following sections. The full implementation of \textsc{BORN} is available in~\cite{mycode}.
	
	\section{Stage 1: Sensing}	\label{sec:stage1}
	This section focuses on developing an algorithm for learning the QM efficiently under a Rademacher feature distribution with theoretical guarantees. In other words, given a dataset $\mathcal{D}$, the objective is to estimate the model coefficients $(\Mtrue, \wtrue)$.
	\subsection{Operators and Notations}
	For a mini-batch $\mathcal D^{(t)}$ with batch-size $B$, we define the design matrix as $\bm X_t = [\bm x_{1,t},\dots,\bm x_{B,t}]^\top \in \mathbb R^{B\times N}$ and the response vector as $\bm y_t = (y_{1,t},\dots,y_{B,t})\in \mathbb R^{B\times 1}$. Next, for batch $t$, any $\bm M$ and  $\bm y$, we define the linear operator $\mathcal A^{(t)}$ and its adjoint operator $\mathcal A'^{(t)}$ as follows
	\begin{equation}
		\begin{aligned}
			\mathcal A^{(t)}(\bm M) & \triangleq \SquareBr{\bm x_{1,t}^\top \bm M \bm x_{1,t} ,\dots, \bm x_{B,t}^\top \bm M \bm x_{B,t}}, \\
			\mathcal A'^{(t)}(\bm y) & \triangleq \sum_{b=1}^B y_b \bm x_{b,t} \bm x_{b,t}^\top.
		\end{aligned}
	\end{equation}
		For mini-batch $t$, let $(\bm M_t, \bm w_t)$ be the estimate, we define
	\begin{equation} \label{eq:update}
		\begin{aligned}
			\bm M_t &\triangleq\frac{1}{2}\big(\bm U_t \bm V_t^\top  + \bm V_t \bm U_t^\top \big), \\
			\bm H_1^{(t)} &\triangleq \frac{1}{2B}\mathcal{A}'^{(t)}\bigg(\bm y_t - \mathcal A^{(t)}(\bm M_t) - \bm X_t \bm w_t\bigg), \\
			h_2^{(t)} &\triangleq \frac{1}{B} \bm 1^\top \bigg(\bm y_t - \mathcal A^{(t)}(\bm M_t) - \bm X_t \bm w_t\bigg), \\
			\bm h_3^{(t)} &\triangleq \frac{1}{B} \bm X_t^\top  \bigg(\bm y_t - \mathcal A^{(t)}(\bm M_t) - \bm X_t \bm w_t\bigg). \\
		\end{aligned}
	\end{equation}
	Herein, we parameterize $\bm M_t$ via the standard low-rank factorization, where $\bm U_t,\bm V_t\in\mathbb{R}^{N\times r}$ are low-rank factor matrices. In addition, $\bm H_1^{(t)}, h_2^{(t)}$ and $\bm{h}_3^{(t)}$ are simple feedback quantities that summarize the mismatch between the measurements and the current estimate. Intuitively, $\bm H_1^{(t)}$ represents the update of the quadratic term, $h_2^{(t)}$ models the mean offset of the mismatch, and $\bm{h}_3^{(t)}$ represents the update of the linear term.
	\subsection{Matrix Sensing Algorithm}
	Our method is inspired by prior work on symmetric matrix sensing \cite{chen2016solvingrandomquadraticsystems, Lin2016gFM} to estimate the model parameters. These studies primarily focus on Gaussian designs and do not directly cover the Rademacher designs induced by binary RIS. Hence, while the main algorithm can be adopted, the theoretical analysis is fundamentally different, and we provide dedicated theoretical guarantees for the Rademacher case in Section~\ref{sec:theoretical_analysis}. The pseudo-code of $\mathtt{MatrixSensing}$ algorithm is given in Algorithm \ref{alg: gFM}, and we explain the details below.
	\begin{algorithm}[t]
		\caption{$\mathtt{MatrixSensing}$ }
		\begin{algorithmic}[1]\label{alg: gFM}   
			\STATE \textbf{Input:} Collection of data $\mathcal D$.
			\STATE \textbf{Output:} $\Mest, \west$.	
			\STATE \textbf{Parameters:} Batch-size $B$, number of epochs $T$, rank $r$.
			\STATE Initialize: $\bm w_0 =\bm 0_{N}$, $\bm V_0 = \bm 0_{N\times r}$, \\ 
				\quad \quad \quad \quad \! $\bm U_0 = \mathrm{SVD}(\bm H_1^{(0)} - \frac{1}{2}h_2^{(0)}\bm I, r)$, $\bm M_0=\bm 0_{N\times N}$.
			\FOR{$t = 0, 1, \cdots, T-1$}
			\STATE Draw the mini-batch $\{\bm{X}_{t},\bm{y}_{t}\}$ from $\mathcal D$.
			\STATE Compute $\bm{\widehat U}_{t+1} =  (\bm H_1^{(t)\top} - h_2^{(t)} \bm I + \bm M_{t}) \ \bm U_{t}$.
			\STATE Orthogonalize $\bm{\widehat U}_{t+1}$ as $\bm U_{t+1} = \mathrm{QR}(\bm{\widehat U}_{t+1})$.
			\STATE Compute $\bm V_{t+1} = (\bm H_1^{(t)\top} - h_2^{(t)} \bm I + \bm M_{t}) \bm U_{t+1}$.
			\STATE Compute $\bm M_{t+1}$ using \eqref{eq:update}.
			\STATE Compute $\bm w_{t+1} = \bm h_3^{(t)} + \bm w_{t}$.
			\ENDFOR    
			\STATE Compute $\Mest = \Pi_{\mathcal S_+^N, r}(\bm M_T)$ and $\west = \bm w_T$.
			\label{line: projection}
			\STATE \textbf{return} $\Mest, \west$.	
		\end{algorithmic}
	\end{algorithm}
	
	The data $\mathcal{D}$ is partitioned into a number of mini-batches, each consisting of $B$ data samples. In each batch, $\mathtt{MatrixSensing}$ processes $B$ training instances and alternates between parameter updates. This procedure continues for $T$ mini-batch iterations. Intuitively, at each round $t$, the algorithm attempts to estimate the errors $(\Mtrue - \bm M_t)$ and $(\wtrue - \bm w_t)$ via
	\begin{equation}
		\begin{aligned}
			\Mtrue - \bm M_t \approx \bm H_1^{(t)} - h_2^{(t)}\bm I; \quad \wtrue -\bm w_t \approx \bm h_3^{(t)} .
		\end{aligned}
	\end{equation}
	Using these error estimates, the algorithm converges at a linear rate, as we will show in Section \ref{sec:theoretical_analysis}. 
	\vspace{-0.4cm}
	\subsection{Matrix Projection}
	It is important to emphasize that the final step of Algorithm~\ref{alg: gFM} (line~\ref{line: projection}) enforces the output $\Mest$ to be a PSD matrix. If the sole objective were parameter estimation, one could directly return $(\bm M_T,\bm w_T)$. However, even with arbitrarily small estimation error, there is no guarantee that $\bm M_T$ is PSD. Without this property, the subsequent optimization stage would face severe computational difficulties. To overcome this issue, we introduce a projection step that ensures $\Mest$ lies in the PSD cone. Owing to this, the optimization stage can be executed efficiently in linear time, which will be shown in Section \ref{sec:optimization}.
	
	To explain, let $\mathcal S_+^N \subset \mathbb R^{N\times N}$ denote the space of PSD matrices, and let $\Pi_{\mathcal S_+^N,r}$ be the projection operator onto the set of PSD matrices with rank at most $r$, under the Frobenius norm. This projection can be computed via truncated eigenvalue decomposition, as summarized in the following proposition.
	\begin{proposition} \label{prop: Projection to PSD matrices}
		Let $\bm M \in \mathbb R^{N\times N}$ be a symmetric matrix. 
		Let $\bm M = \bm Q^\top \bm D\bm Q$ be eigen-decomposition of $\bm M$, where $\bm D =\Diag{\lambda_1,...,\lambda_N}$ such that $\lambda_1\geq\lambda_2\geq...\geq \lambda_N$.
		Let $\lambda_i^+ = \max\{0, \lambda_i\}$, then define $\bm D^+ =  \Diag{[\lambda_1^+,..., \lambda_r^+, \bm 0_{N-r}]}$.
		Let $\bm M^+ = \bm Q^\top \bm D^+ \bm Q$. We have 
		\begin{equation}
			\bm M^+ = \Pi_{\mathcal S_+^N, r}(\bm M) \triangleq \argmin_{\substack{\bm M'  \in \mathcal S_+^N \\ \Rank{\bm M'} \leq r} } \| \bm M' - \bm M\|_F.
		\end{equation}
	\end{proposition}
	\begin{proof}
		See Appendix \ref{proof: Projection to PSD matrices}.
	\end{proof}
	After this step, Algorithm \ref{alg: gFM} outputs $(\Mest,\west)$, which not only approximates the true parameters but also ensures that $\Mest$ is PSD with rank at most $r$, which is essential for the optimization stage.
	
	\begin{remark}[\textbf{Computational Complexity of Algorithm \ref{alg: gFM}}] \label{re:computational_complexity_alg_1}
		The truncated $\mathrm{SVD}$ and $\mathrm{QR}$ decomposition cost $O(r^{2}N)$ each, while multiplying an $N\times N$ by an $N\times r$ matrix costs $O(N^{2}r)$. The adjoint operator costs $O(BN^{2})$. With $r \le 2$, the dominant term is $O(B N^{2})$, hence, the total complexity is $O(T B N^{2})$.
	\end{remark}
	\section{Stage 2: Optimization}
	\label{sec:optimization}
	\begin{algorithm}[t]
		\caption{$\mathtt{QuadraticOptimizing}$} \label{alg:QuadraticOptimizing}
		\begin{algorithmic}[1]
			\STATE \textbf{Input}: $\Mest, \west$ obtained from sensing stage.
			\STATE \textbf{Output}: $\widehat{\bm{x}}$.
			\STATE Compute $\breve{\bm V}$ and $ \breve{\bm v}_0$ such that \\
			\quad \quad $\Mest =  \breve{\bm V}^\top  \breve{\bm V}, \quad  \breve{\bm v}_0 = \frac{1}{2} ( \breve{\bm V}^\top)^+ \west$. \label{line:SVD}
			\FOR {$n=0$ to $N$} \label{line:starting_flip}
			\STATE Compute flip points $\bm{a}_n^-$ and $\bm{a}_n^+$ using (\ref{eq:changing_points}).
			\ENDFOR \label{line:ending_flip}
			\FOR {$m=1$ to $2N+2$}
			\STATE Compute $\bm{w}_m$.
			\STATE Let $\widehat{\bm{a}}_m = \frac{\bm{w}_m }{\|\bm{w}_m\|}$ if it lies in the interior of the $m$-th arc, otherwise $\widehat{\bm{a}}_m$ equals either the endpoint of the $m$-th arc yielding larger $g(\widehat{\bm{a}}_m)$. \label{line:select_am}
			\ENDFOR
			\STATE $\widehat{\bm{a}} = \widehat{\bm{a}}_s$ where $s = \arg \underset{m}{\max} \ g(\widehat{\bm{a}}_m)$.
			\FOR {$n=1$ to $N$}
			\STATE $\widehat{x}_n = \operatorname{sgn}(\breve{\bm v}_n^\top \widehat{\bm{a}}) \cdot \operatorname{sgn}(\breve{\bm v}_0^\top \widehat{\bm{a}})$ (set to $+1$ if product is $0$).
			\ENDFOR
			\RETURN $\widehat{\bm{x}} = [\widehat{x}_1, \dots, \widehat{x}_N]^\top$.
		\end{algorithmic}
	\end{algorithm}
	
	This section describes the optimization stage, focusing on finding $\widehat{\bm{x}}$ given the system parameters $(\Mest, \west)$ achieved in the sensing stage. In other words, we aim to solve the following optimization problem:
	\begin{equation}
		\begin{aligned} \label{eq:optimization_with_known_parameters}
			\mathrm{(P2)} \quad \underset{\bm{x}}{\mathrm{maximize}} \quad & \widehat{y}(\bm{x}) = \bm{x}^\top \Mest \bm{x} + \bm{x}^\top \west, \\
			\mathrm{subject \ to} \quad & \bm{x} \in \{ -1, 1\}^N.
		\end{aligned}
	\end{equation}
	This is a quadratic optimization over the binary domain $\{\pm 1\}^N$, which is generally challenging. In general, without the PSD property of $\Mest$, such problems are NP-hard. In principle, common methods such as manifold optimization with projection, branch-and-bound, or semidefinite relaxation (SDR) can be applied. Nevertheless, manifold-based methods are iterative and generally provide convergence only to stationary points, and the final projection to the binary set may further incur performance loss. Branch-and-bound has exponential worst-case complexity, while SDR becomes computationally heavy as $N$ grows.
	
	In our case, after the PSD projection in Stage~1, $\Mest$ is PSD and low-rank. Let $\hat r \triangleq \operatorname{rank}(\Mest)$. We first reformulate $\mathrm{(P2)}$ in a form that is valid for any $\hat r \ge 1$, and then develop the corresponding optimization method according to the value of $\hat r$. In the single-user setting, which is the focus of this paper, Proposition~\ref{prop: Projection to PSD matrices} implies that $\hat r \le 2$. In this case, the nontrivial generic case is $\hat r=2$, for which $\mathrm{(P2)}$ can be solved \emph{globally} in $O(N)$. In the multi-user extension, typically $\hat r > 2$, and we therefore propose a low-complexity alternating algorithm that yields a monotone improvement of the objective.
	
	To start, recall that $\Mest$ is PSD, it is feasible to decompose $\Mest = \breve{\bm V}^\top \breve{\bm V}$. 
	The objective function now becomes
	\begin{align}
		\widehat{y}(\bm{x}) = \|\breve{\bm V} \bm{x} + \breve{\bm v}_0\|^2 - \|\breve{\bm v}_0\|^2,
	\end{align}
	where $\breve{\bm v}_0 = \frac{1}{2} (\breve{\bm V}^\top)^+ \west.$ 
	Since $\|\breve{\bm v}_0\|^2$ is a constant independent of $\bm{x}$, maximizing $\widehat{y}(\bm{x})$ is equivalent to maximizing $\|\breve{\bm V} \bm{x} + \breve{\bm v}_0\|^2$. 
	Let $ \breve{\bm v}_n$ denote the $n$-th column of $ \breve{\bm V}$. By introducing an auxiliary unit vector $\bm{a} \in \mathbb{R}^{\hat r}$ with $\|\bm{a}\|=1$ and noting that $\|\bm{z}\|^2 = \underset{\|\bm{a}\|=1}{\max} (\bm{z}^\top \bm{a})^2$ for any $\bm{z}$, our optimization problem is equivalent to
	\begin{equation}
		\max_{ \bm x \in \{\pm 1\}^N} \max_{\|\bm{a}\|=1} F(\bm a, \bm x),
	\end{equation}
	where $F(\bm a, \bm x) \triangleq | (\breve{\bm V}\bm{x} + \breve{\bm v}_0)^\top \bm a| = \big| \breve{\bm v}_0^\top \bm{a} + \sum_{n=1}^N x_n ( \breve{\bm v}_n^\top \bm{a}) \big|$.
	For a fixed $\bm{a}$, to maximize the magnitude, every term $x_n \breve{\bm v}_n^\top \bm{a}$ should have the same sign with $\breve{\bm v}_0^\top \bm{a}$. In other words, the optimal $\bm x$ for a fixed $\bm{a}$ is
	\begin{align}
		x_n^\star = \mathrm{sgn}( \breve{\bm v}_0^\top \bm{a} \cdot  \breve{\bm v}_n^\top \bm{a}).
		\label{eq:fix_a_find_x}
	\end{align}
	\vspace{-0.8cm}
	\subsection{$\hat r = 2:$ Single-user Setting:}
	In the single-user setting considered in this paper, we typically have that $\hat r = 2$ thanks to Proposition \ref{prop: Projection to PSD matrices}. Hence, $\bm a$ lies on the unit circle. We can obtain the global optimum efficiently in $O(N)$ time \cite{9779545} using $\mathtt{QuadraticOptimizing}$ algorithm, shown in Algorithm \ref{alg:QuadraticOptimizing}. To explain, substituting \eqref{eq:fix_a_find_x} into $F(\bm a,\bm x)$ yields the equivalent optimization as
	\begin{align}
		\max_{\|\bm{a}\|=1} g(\bm{a}) = | \breve{\bm v}_0^\top \bm{a}| + | \breve{\bm v}_1^\top \bm{a}| + \cdots + | \breve{\bm v}_N^\top \bm{a}|.
	\end{align}
	The core idea is to partition the unit circle into arcs where the signs of $ \breve{\bm v}_n^\top \bm{a}$ remain constant within each arc. Considering the term $ \breve{\bm v}_n^\top \bm{a}$ and let $\psi_n$ be the angle of $ \breve{\bm v}_n$, the sign flips at two flip points satisfying $ \breve{\bm v}_n^\top \bm{a} = 0$, which are
	\begin{equation}
		\begin{aligned}
			\label{eq:changing_points}
			\bm{a}_n^- &\triangleq [\cos(\psi_n - \pi/2), \sin(\psi_n - \pi/2)]^\top, \\
			\bm{a}_n^+ & \triangleq [\cos(\psi_n + \pi/2), \sin(\psi_n + \pi/2)]^\top.
		\end{aligned}
	\end{equation}
	The set $\{\bm{a}_0^-, \bm{a}_0^+, \dots, \bm{a}_N^-, \bm{a}_N^+\}$ divides the circle into at most $2N+2$ arcs. Within the $m$-th arc, since the sign of each term $ \breve{\bm v}_n^\top \bm{a}$ does not change, $g(\bm{a})$ simplifies to a linear form
	\begin{align}
		g(\bm{a}) = \sum_{ \breve{\bm v}_n \in \mathcal{L}^+} ( \breve{\bm v}_n^\top \bm{a}) + \sum_{\breve{\bm{v}}_n \in \mathcal{L}^-} (- \breve{\bm v}_n^\top \bm{a}) = \bm{w}_m^\top \bm{a},
	\end{align}
	where $\mathcal{L}^+ = \{ \breve{\bm v}_n \mid  \breve{\bm v}_n^\top \bm{a} \geq 0\}, \mathcal{L}^- = \{ \breve{\bm v}_n \mid  \breve{\bm v}_n^\top \bm{a} < 0\} $ and
	\begin{align}
		\bm{w}_m = \!\! \sum_{ \breve{\bm v}_n \in \mathcal{L}^+} \breve{\bm v}_n - \sum_{ \breve{\bm v}_n \in \mathcal{L}^-}  \breve{\bm v}_n, \ \ \text{ for } m \in [2N + 2].
	\end{align}
	The maximum over each arc can be efficiently computed as described in line \ref{line:select_am}, Algorithm \ref{alg:QuadraticOptimizing}.
	\begin{remark}(\textbf{Computational Complexity of Algorithm \ref{alg:QuadraticOptimizing}}) \label{re:computational_complexity_alg_2}
		Constructing $\breve{\bm V}$ and $\breve{\bm v}_0$ (line \ref{line:SVD}) costs $O(r^2 N)$ operations, while forming $2N$ flip points (shown in line \ref{line:starting_flip} - \ref{line:ending_flip}) needs $O(N)$ operations. Updating $\bm{w}_m$ needs $O(1)$ per arc. Hence, the total computational complexity is $O(N)$.
	\end{remark}
	\vspace{-0.4cm}
	\subsection{$\hat r > 2$:}
	When $\hat r>2$, $\bm a$ lies on the unit sphere and obtaining the global optimum is no longer straightforward, since the sign-change boundaries $\breve{\bm v}_n^\top \bm a=0$ partition the sphere into many cells. Therefore, instead of exhaustive search over all sign regions, we adopt a low-complexity alternating algorithm. Specifically, for a fixed $\bm a$, the optimal $\bm x$ is already given in \eqref{eq:fix_a_find_x}. On the other hand, for a fixed $\bm x$, maximizing $F(\bm a,\bm x)$ over all unit vectors $\bm a$ is equivalent to solving ${\max} \left| \bm a^\top (\breve{\bm V}\bm x+\breve{\bm v}_0)\right|,$ subject to $\|\bm a\|=1$. By the Cauchy-Schwarz inequality, the optimum is attained at
	\begin{align}
		\bm a^\star=\frac{\breve{\bm V}\bm x+\breve{\bm v}_0}{\|\breve{\bm V}\bm x+\breve{\bm v}_0\|}.
	\end{align}
	Based on these closed-form updates, we alternately optimize $\bm x$ and $\bm a$ until convergence. The pseudo-code is summarized in Algorithm~\ref{alg:alt_ascent}.
	
	This algorithm yields a monotone ascent of the objective $F(\bm a,\bm x)$. Indeed, for fixed
	$\bm a^{(k)}$, the update $\bm x^{(k+1)}$ maximizes $F(\bm a^{(k)},\bm x)$, while for fixed
	$\bm x^{(k+1)}$, the update $\bm a^{(k+1)}$ maximizes $F(\bm a,\bm x^{(k+1)})$. Hence,
	\begin{align}
		F(\bm a^{(k)},\bm x^{(k)}) \le F(\bm a^{(k)},\bm x^{(k+1)}) \le F(\bm a^{(k+1)},\bm x^{(k+1)}). \nonumber
	\end{align}
	Therefore, the objective sequence is nondecreasing and convergent since $F(\bm a,\bm x)$ is bounded above.
	\begin{remark}(\textbf{Computational Complexity of Algorithm \ref{alg:alt_ascent}}).
		Each iteration requires computing $N$ inner products $\breve{\bm v}_n^\top \bm a$ and one matrix-vector
		product $\breve{\bm V}\bm x$. Hence, the per-iteration complexity is $O(\hat{r} N)$.
	\end{remark} 
	\vspace{-0.3cm}
	\subsection{Extension to Multi-user Setting:}
	We briefly discuss an extension to a multi-user setting with $K_u$ single-antenna users sharing the same RIS configuration. For user $k$, by the same procedure as in Section~\ref{sec:problem_formulation}, the received SNR can be written as $s_k(\bm x) = \bm x^\top \bm M_k \bm x + \bm x^\top \bm w_k + c_k ,$ where $\bm M_k \succeq \bm 0$ and $\operatorname{rank}(\bm M_k)\leq 2$. A common flexible design criterion is the weighted sum SNR, formulated as
	\begin{equation}
		\begin{aligned}
			\mathrm{(P3)} \quad \underset{\bm x}{\mathrm{maximize}}
			\quad &
			\sum_{k=1}^{K_u} \beta_k s_k(\bm x) \\
			\mathrm{subject\ to}
			\quad &
			\bm x \in \{-1,1\}^{N},
		\end{aligned}
	\end{equation}
	where $\beta_k \ge 0$ denotes the priority weight for user $k$. This problem is equivalent to
	\begin{equation}
		\begin{aligned}
			\underset{\bm x \in \{ \pm 1 \}^{N}}{\mathrm{maximize}} \quad & \bm x^\top \bar{\bm M} \bm x + \bm x^\top \bar{\bm w},
		\end{aligned}
	\end{equation}
	where $\bar{\bm M} = \sum_{k=1}^{K_u} \beta_k \bm M_k$ and $\bar{\bm w} = \sum_{k=1}^{K_u} \beta_k \bm w_k.$ Since $\bm M_k$ is PSD, $\bar{\bm M}$ is also PSD. It is important to note that now $\Rank{\bar{\bm M}}$ is typically larger than 2. Therefore, the multi-user case generally falls into the $\hat r>2$ regime.
	
	In the \textit{sensing stage}, the RIS still tries out the same random configurations as before, while each user records its own SNR measurements. The same sensing procedure is then applied separately to estimate $(\bm M_k,\bm w_k)$ for all $k$, after which $\bar{\bm M}$ and $\bar{\bm w}$ are formed by weighted summation. In the \textit{optimization stage}, we simply replace $(\Mest,\west)$ in $\mathrm{(P2)}$ by $(\bar{\bm M},\bar{\bm w})$. Since the resulting problem has the same form as $\mathrm{(P2)}$ but typically with $\hat r>2$, it can be solved using Algorithm~\ref{alg:alt_ascent}.
	\begin{algorithm}[t]
		\caption{$\mathtt{AlternatingOptimization}$}
		\label{alg:alt_ascent}
		\begin{algorithmic}[1]
			\STATE \textbf{Input}: $\Mest, \west$ obtained from sensing stage.
			\STATE \textbf{Output}: $\widehat{\bm{x}}$.
			\STATE Compute $\breve{\bm V}$ and $\breve{\bm v}_0$, similar to step 3, Algorithm \ref{alg:QuadraticOptimizing}.
			\STATE Initialize $\bm a^{(0)}$ such that $\|\bm a^{(0)}\|=1$, and set $k=0$.
			\REPEAT
			\FOR{$n=1$ to $N$}
			\STATE $x_n^{(k+1)}=\mathrm{sgn}\!\left(\breve{\bm v}_0^\top \bm a^{(k)} \cdot \breve{\bm v}_n^\top \bm a^{(k)}\right)$ (set to $+1$ if the product is $0$).
			\ENDFOR
			\STATE $\bm a^{(k+1)}=\dfrac{\breve{\bm V}\bm x^{(k+1)}+\breve{\bm v}_0}
			{\|\breve{\bm V}\bm x^{(k+1)}+\breve{\bm v}_0\|}$.
			\STATE $k \leftarrow k+1$.
			\UNTIL{$\bm x^{(k)}=\bm x^{(k-1)}$}.
			\STATE \textbf{return} $\widehat{\bm x}=\bm x^{(k)}$.
		\end{algorithmic}
	\end{algorithm}
	
	\section{Theoretical Analysis}
	\label{sec:theoretical_analysis}
	In this section, we provide a theoretical guarantee on the suboptimality gap of $\widehat{\bm x}$ produced by \textsc{BORN}. Before presenting the main result, we first state and explain the assumptions required for the success of the algorithm.
	\vspace{-0.3cm}
	\subsection{Assumption} \label{subsec: Assumption}
	We note that theoretical guarantees cannot be established in a general case. The main challenge arises in the sensing stage, where Algorithm~\ref{alg: gFM} relies on restrictive conditions on higher-order moments of the distribution of $\bm x$, which does not hold for arbitrary probability distributions over $\{-1,1\}^N$ \cite{lin2017second,Lin2016gFM}. Without additional assumptions or special structure on the coefficients, when $\bm x$ is sampled from $\{-1,1\}^N$, the diagonal entries of $\Mtrue$ are not recoverable in general \cite{cai2015rop,lin2017second}. A canonical example is when $\Mtrue$ is a diagonal matrix. In this case, all off-diagonal entries are zero and provide no information on the diagonal. Leveraging this fact, we introduce the following incoherence assumption, which rules out this extreme case and ensures that the diagonal can be inferred from the off-diagonal entries.
	
	\begin{definition}[\textbf{Incoherence constant}]
		Let $\bm M \in \mathbb R^{N \times N}$ be a symmetric matrix of rank $r$ with eigenvalue decomposition $\bm M = \bm U \bm \Sigma \bm U^\top$.
		We say matrix $\bm M$ satisfies the incoherence assumption with positive constant $\mu_0(\bm M)$ if and only if for any $i \in [N]$,
		\begin{equation} \label{eq: def of Incoherence}
			\|\bm U_{i,:} \|^2_2 \leq \frac{\mu_0(\bm M) r }{N}.
		\end{equation}
	\end{definition}
	
	\begin{assumption}[\textbf{Boundedness of incoherence constant}] \label{Assp: Incoherence}
		Considering the sequence $(\bm M_t)_{t=1}^T$ computed by Algorithm \ref{alg: gFM}, define
		$\bar \mu_0 \triangleq \underset{t \in [T]}{\max} \ \mu_0(\Mtrue - \bm M_t).$
		We assume that
		\begin{equation} \label{eq: Incoherence assumption}
			\bar \mu_0 = O\left( \frac{N\sigma_r(\Mtrue)}{r(\|\Mtrue \|_2 + \|\wtrue \|_2)} \right).
		\end{equation}
	\end{assumption}
	
	Assumption \ref{Assp: Incoherence}, also known as the incoherence assumption in the matrix completion literature, is crucial for the success of various matrix completion algorithms \cite{Chi2019_MatrixCompletionIncoherence}. In our setting, it ensures that small errors in estimating the off-diagonal entries of $(\bm M_t - \Mtrue)$  lead to a small spectral error $\|\bm M_t - \Mtrue\|_2$.
	
	Regarding the constant $\bar\mu_0$, we note that when $N \gg r$, the incoherence measure $\mu_0(\bm M_t - \Mtrue)$ in \eqref{eq: def of Incoherence} is typically small, and hence so is $\bar\mu_0$. Furthermore, since the right-hand side (\textsc{RHS}) of \eqref{eq: Incoherence assumption} grows linearly with $N$, Assumption \ref{Assp: Incoherence} is very likely to hold when $N$ is sufficiently large. This phenomenon can be viewed as a blessing of dimensionality and is further verified in our experiments in Section \ref{sec:experiments}.
	
	\subsection{The Analysis of Suboptimality Gap}
	Our main result, the suboptimality gap of \textsc{BORN}, is introduced in the following theorem.
	\begin{theorem} [\textbf{Suboptimality Gap of \textsc{BORN}}]\label{theorem: main optimal gap}
		Suppose Assumption \ref{Assp: Incoherence} holds.
		Regarding Algorithm \ref{alg: gFM}, choose the batch-size and number of epochs as follows 
		\begin{equation} \label{eq: BORN parameters}
			\begin{aligned}
				B &= \frac{Cr^3 N\log_2(N\eta^{-1})}{\delta^2} \text{ with }
				\delta = \frac{C\sigma_r(\Mtrue)}{\|\Mtrue\|_2 + \|\wtrue\|_2}, \\
				T &=  \log_2(14Nr(\|\Mtrue\|_2 + \|\wtrue\|_2)\varepsilon^{-1}).
			\end{aligned}
		\end{equation}
		Let $\xtrue$ and $\widehat{\bm x}$ be the optimal solution and output of \textsc{BORN}, respectively. Then, with probability at least $1-\eta$,
		\begin{equation}
			y(\bm x_\star)  - y(\widehat{\bm x}) \leq \varepsilon.
		\end{equation}
	\end{theorem}
	\begin{proof}
		See Appendix~\ref{proof of theorem optimal gap}.
	\end{proof}
	\begin{remark}[\textbf{On suboptimality gap}]
		To the best of our knowledge, Theorem \ref{theorem: main optimal gap} provides the first result that establishes a provable suboptimality gap using only $O(N \log_2(N))$ samples, which is near-linear in $N$. To compare, existing algorithms such as \textsc{RFocus}~\cite{arun2020rfocus}, \textsc{CSM}~\cite{9940989} and \textsc{GCSM}~\cite{10694808} do not provide such finite-sample suboptimality guarantees.
		Therefore, \textsc{BORN} stands out as the only algorithm that provides a suboptimality gap with good sample complexity.
	\end{remark}
	
	\begin{remark}[\textbf{On computational complexity}]
		Moreover, from Remark~\ref{re:computational_complexity_alg_1} and \ref{re:computational_complexity_alg_2}, \textsc{BORN} runs in at most  $O(r^3 N^2 \log_2(N/\epsilon))$ time with the specific choices of $B$ and $T$ in Theorem \ref{theorem: main optimal gap}.
		This, combined with its strong performance guarantees, makes \textsc{BORN} both statistically and computationally efficient.
	\end{remark}
	\subsection{Supporting Analysis}
	We dedicate this subsection to outlining the proof of Theorem \ref{theorem: main optimal gap} and discussing its technical challenges.
	The proof of Theorem~\ref{theorem: main optimal gap} proceeds in three steps:
	(1) The sensing stage produces estimates with small error, as demonstrated in Theorem~\ref{theorem: main estimation error}.
	(2) The optimization stage achieves an optimal solution with respect to the estimated model, as shown in Lemma~\ref{lemma:suboptimality_gap_optimization}.
	(3) The small estimation error from the sensing stage leads to a small suboptimality gap, as established in Lemma~\ref{lemma: estimation error to optimal gap}.
	
	With this outline in place, we now establish the estimation error bound of Algorithm \ref{alg: gFM}, which serves as a key component in the analysis of the suboptimality gap.
	In particular, the next theorem shows that, even under the Rademacher distribution, $\Mtrue$ and $\wtrue$ are learnable as long as the model coefficient matrix satisfies the incoherence property.
	We note that this result is nontrivial, since theoretical guarantees are generally impossible to obtain under Rademacher distribution \cite{lin2017second}.
	\begin{theorem}[\textbf{Estimation Error of Algorithm 1}] \label{theorem: main estimation error}
		Suppose Assumption \ref{Assp: Incoherence} holds. Regarding Algorithm \ref{alg: gFM}, choose the batch-size and number of epochs as follows
		\begin{equation} \label{eq: gFM parameters}
			\begin{aligned}
				B &= \frac{Cr^3 N\log_2(N\eta^{-1})}{\delta^2} \text{ with }
				\delta = \frac{C\sigma_r(\Mtrue)}{\|\Mtrue\|_2 + \|\wtrue\|_2}, \\
				T &= \log_2((\|\Mtrue\|_2 + \|\wtrue\|_2)\varepsilon^{-1}),
			\end{aligned}
		\end{equation}
		where $C>0$ is a universal constant, $\varepsilon \in(0,1]$ denotes an estimation error, and $\eta\in(0,1)$. Then, we have that $\Mest$ is a symmetric PSD matrix of rank at most $r$, and with probability at least $1-\eta$,
		\begin{equation}
			\begin{aligned}
				\|\Mest - \Mtrue\|_2 \leq 6r\varepsilon, \quad \quad \|\west - \wtrue\|_2 \leq \varepsilon.
			\end{aligned}
		\end{equation}
	\end{theorem}
	\begin{proof}
		See Appendix~\ref{proof of theorem main estimation error}.
	\end{proof}
	Regarding the optimization stage, we show that Algorithm \ref{alg:QuadraticOptimizing} achieves a globally optimal solution with respect to the estimated model; hence, no additional suboptimality is incurred during the optimization stage.
	\begin{lemma} \label{lemma:suboptimality_gap_optimization}
		Let $\widehat{\bm x}$ be the output of Algorithm \ref{alg:QuadraticOptimizing} with input $(\Mest, \west)$. Then, $\widehat{\bm x}$ satisfies: 
		$\widehat{\bm x}\in \underset{\bm x\in \{ \pm 1\}^N}{\argmax} \ \bm x^\top \Mest \bm x + \bm x^\top \west.$
	\end{lemma}
	\begin{proof}
		In Algorithm \ref{alg:QuadraticOptimizing}, all steps are order-preserving reformulations of the objective and the search is performed in a finite candidate set, hence, $\widehat{\bm x}$ is globally optimal.
	\end{proof}
	Finally, we show that a small estimation error indeed leads to a small suboptimality gap.
	\begin{lemma} \label{lemma: estimation error to optimal gap}
		Suppose the estimation error is bounded as, 
		\begin{align*}
			\| \Mest - \Mtrue \|_2 \leq \epsilon_M, \quad \quad \quad  \| \west - \wtrue \|_2 \leq \epsilon_w.
		\end{align*}
		Then, the suboptimality gap of $\xest$ can be bounded as:
		\begin{equation}
			y(\xtrue) - y(\xest) \leq 2N\epsilon_M + 2\sqrt{N} \epsilon_w.
		\end{equation}
	\end{lemma} 
	\begin{proof}
		See Appendix \ref{proof of lemma estimation error to optimal gap}
	\end{proof}
	\begin{figure*}[t]
		\centering
		\includegraphics[width=1\textwidth]{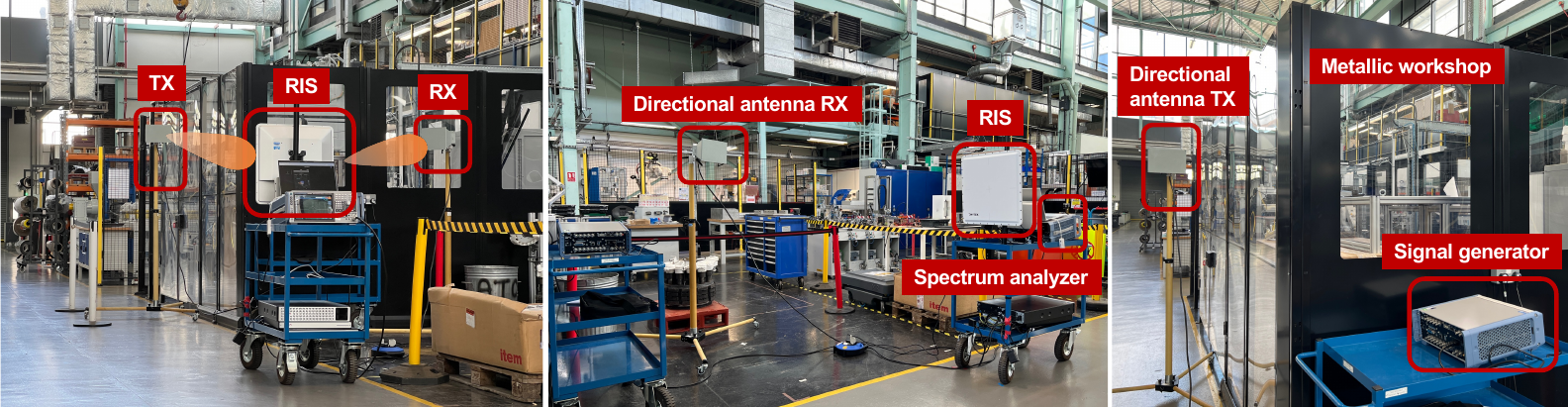}
		\caption{Our field test setup at 4.7~GHz in a WMG International Manufacturing Centre building.}
		\label{fig:measurements}
	\end{figure*}
	\section{Experiments}
	\label{sec:experiments}
	In this section, we benchmark the performance of \textsc{BORN} against existing blind beamforming algorithms, as well as the optimal benchmark, through both field tests and simulations. CSI-assisted, learning approaches typically require coherent pilot feedback or offline training datasets. A fully fair comparison would therefore require redefining the sensing overhead and protocol assumptions, and is left outside the scope of this work. A total of 6 algorithms are considered:
	\begin{itemize}
		\item \textsc{BORN}: Our proposed algorithm. In our implementation, the dataset $\mathcal{D}$ is shuffled and reused multiple times. The full implementation is available at~\cite{mycode}.
		\item \textsc{RFocus}~\cite{arun2020rfocus}: One of the earliest blind beamforming algorithms, which relies on a majority-voting mechanism.  
		\item \textsc{CSM}~\cite{9940989}: The Conditional Sample Mean algorithm, which configures RIS elements based on conditional averages of measurement data.  
		\item \textsc{GCSM}~\cite{10694808}: An extension of \textsc{CSM}, where RIS elements are grouped to construct a virtual LOS path.  
		\item Random Max Sampling (\textsc{RMS}): An algorithm that selects the configuration $\bm{x}$ which obtains the maximum received SNR from the dataset, i.e., $\bm{x} = \underset{\bm{x} \in \mathcal{D}}{\argmax} \ s(\bm{x})$.  
		\item \textit{Optimal solution}: The global optimum of $\mathrm{(P0)}$ under the same binary constraint. This is computed using perfect CSI and only available in simulations.
	\end{itemize}
	\begin{figure}[t]
		\centering
		\includegraphics[width=0.45\textwidth]{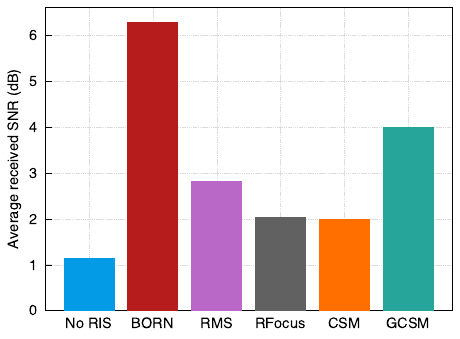}
		\caption{The average received SNR during the field test conducted in a manufacturing building.}
		\label{fig:measurement_result}
	\end{figure}

	\subsection{Field Tests}
	Our field test illustrates the use of RIS and blind beamforming algorithms in a representative manufacturing use case. The measurement campaign was conducted in the International Manufacturing Centre (IMC) building at Warwick Manufacturing Group (WMG), the University of Warwick, the United Kingdom. The test area resembles a real industrial environment. Measurements were taken during a regular working day, when workers were actively moving and transporting goods.  
	
	The TX employed a directional antenna connected to a Rohde \& Schwarz SMBV100B vector signal generator and was placed beside a large metallic workshop. The RX used a directional antenna connected to a Rohde \& Schwarz FSVA3007 spectrum analyzer, positioned in the area adjacent to the workshop. The metallic structure of the workshop blocked the LOS path between TX and RX, as illustrated in Fig.~\ref{fig:measurements}. To overcome this blockage, a RIS~\cite{TMYTEK2025} with 100 electronically switchable elements was deployed approximately $5$~m from the TX and $3$~m from the RX. Since this study focuses on blind beamforming, explicit knowledge of the TX, RX, and RIS coordinates is \textit{not required}. Their positions remained fixed throughout the entire campaign, aligned with a low-dynamics setting to maintain approximate channel stationarity during data collection. The center frequency of the field test is $4.7$~GHz, the transmit power was set to $0$~dBm, and the noise was measured at $-72.38$~dBm prior to data collection. The power of background channel was weak, fluctuating around $-71.23$~dBm including noise power. A total of $2000$ samples were collected during the measurements, with each sample averaged over a period of $2$~seconds. Since the locations of the TX, RX, and RIS were fixed during the field test, our experiment mainly validates \textsc{BORN} in the slow-timescale or quasi-static RIS control regime, where the learned model captures effective channel behavior over the sensing window rather than tracking fast channel variations.

	Fig.~\ref{fig:measurement_result} shows the average received SNR obtained during the field test under various schemes. Without the RIS, the average received SNR is slightly above 1~dB. This very weak baseline is mainly due to the harsh propagation conditions in the factory setting \cite{9971737}. Such conditions lead to an unstable and unreliable background channel, which is representative of many practical industrial environments. With RIS, a naive strategy such as \textsc{RMS} improves the received SNR to about $2.8$~dB, although the fluctuations caused by the dynamic environment remain noticeable. \textsc{RFocus} and \textsc{CSM} perform less effectively in this setting, both yielding only about $2$~dB on average. \textsc{GCSM} achieves better performance, with an average of approximately $4$~dB. Among all evaluated methods, \textsc{BORN} achieves the best performance, boosting the average received SNR to nearly $5$~dB compared to the case of no RIS.
	\begin{figure}[t]
		\centering
		\includegraphics[width=0.42\textwidth]{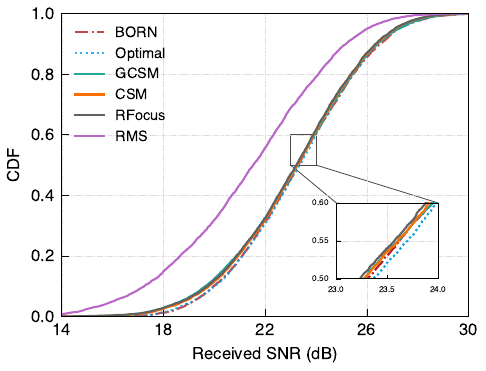}
		\caption{The CDF of received SNR in a LOS condition. In this simulation, $N=60$, $B=600$ and $3000$ samples are used.}
		\label{fig:LOS_N60}
	\end{figure}
	\begin{figure}[t]
		\centering
		\includegraphics[width=0.42\textwidth]{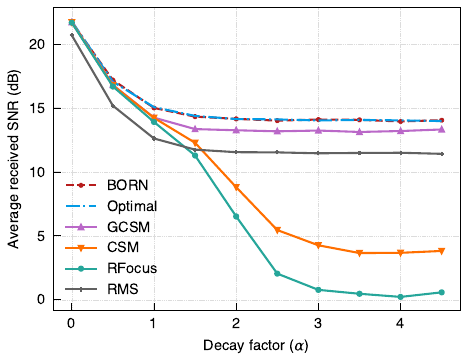}
		\caption{The received SNR when the background channel $h_0$ is attenuated over iterations by a factor of $10^{-\alpha}$. Here, $N=40$, $B=400$ and $N_s = 3000$.}
		\label{fig:decay_h0}
	\end{figure}
	\subsection{Simulation Tests}
	In this section, we benchmark the performance of existing blind beamforming algorithms against the optimal one. We also provide insights related to the behaviors of \textsc{BORN}. The channel model follows the previous studies \cite{9779545}. The background channel $h_0$ is modeled as $h_0 = 10^{-\mathrm{PL}_0 / 20} \xi_0,$ where $\mathrm{PL}_0$ represents the pathloss (in dB) between the TX and RX. This pathloss is computed as $\mathrm{PL}_0 = 32.6 + 36.7 \log(d_0)$, with $ d_0 $ denoting the distance between TX and RX in meters. The Rayleigh fading component $\xi_0$ is drawn from a Gaussian distribution $\mathcal{CN}(0,1)$. Regarding the cascaded reflected channels, $ h_n $ is modeled as 
	\begin{align}
		h_n = 10^{-(\mathrm{PL}_1 + \mathrm{PL}_2) / 20} \xi_{n1} \xi_{n2}, \ \ \  n = 1, 2, \cdots, N,
	\end{align}
	where $\mathrm{PL}_1$ and $\mathrm{PL}_2$ represent the pathlosses for the TX-to-RIS and RIS-to-RX links, respectively. Both are calculated using the pathloss model $\mathrm{PL} \!=\! 30 + 22 \log(d)$, where $ d $ denotes the respective distances in meters. In addition, the Rayleigh fading components $\xi_{n1}$ and $\xi_{n2}$ are independently drawn from $\mathcal{CN}(0,1)$ for each $ n = 1, \ldots, N $.
	
	We use the following parameters unless otherwise stated. The transmit power $P$ is set to $30$ dBm. The background noise power is $\sigma^2 = -90$ dBm. The positions of the TX, RIS, and RX are defined in a 3-dimensional coordinate system, with locations given by $(0, 0, 10)$, $(50, 50, 1.5)$, and $(10, 50, 2)$ in meters, respectively.
		\begin{figure}[t]
		\centering
		\includegraphics[width=0.42\textwidth]{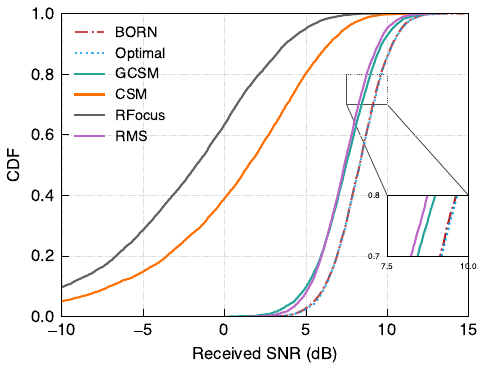}
		\caption{The CDF of received SNR in a NLOS condition. In this simulation, $N=20$, $B=200$ and $1500$ samples are used.}
		\label{fig:NLOS_N20}
	\end{figure}
	\begin{figure}[t]
		\centering
		\includegraphics[width=0.42\textwidth]{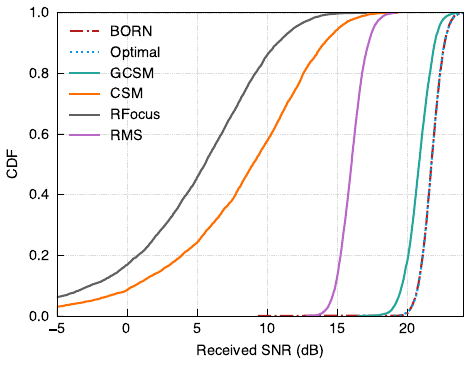}
		\caption{The CDF of received SNR in a NLOS condition. In this simulation, $N=100$, $B=1000$ and $3000$ samples are used.}
		\label{fig:NLOS_N100}
	\end{figure}
	\begin{figure}[t]
		\centering
		\includegraphics[width=0.42\textwidth]{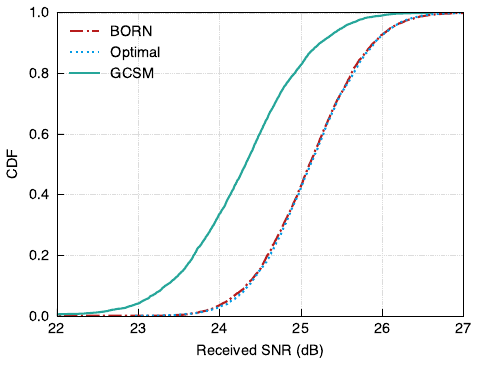}
		\caption{The CDF of received SNR in a NLOS condition. In this simulation, $N=150$, $B=1500$ and $4500$ samples are used.}
		\label{fig:NLOS_N150}
	\end{figure}
	\begin{figure}[t]
		\centering
		\includegraphics[width=0.5\textwidth]{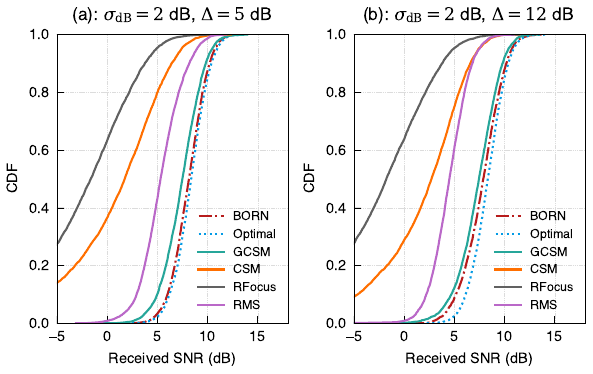}
		\caption{The CDF of received SNR with the quantized and noisy effects on measurements in (a) $\sigma_\mathrm{dB} = 2, \Delta = 5$, and (b): $\sigma_\mathrm{dB} = 2, \Delta = 12$. In these simulations, $N=20$, $B=1000$ and $N_s = 3000$.}
		\label{fig:noisy_quantized_effect}
	\end{figure}
	
	\subsubsection{Performance benchmark in LOS scenarios} Fig.~\ref{fig:LOS_N60} presents the cumulative distribution function (CDF) of the received SNR in a LOS condition. As can be seen, \textsc{BORN} achieves nearly identical performance to the optimal benchmark across the entire distribution, with both reaching approximately $20$~dB at the $10$th percentile. \textsc{CSM} and \textsc{RFocus} attain around $23.7$~dB at the $55$th percentile, while \textsc{GCSM} performs slightly lower at $23.5$~dB. As expected, \textsc{RMS}, being the simplest algorithm, performs the worst with only $18$~dB at the $15$th percentile. Nevertheless, except for \textsc{RMS}, all algorithms perform remarkably well under LOS conditions, despite lacking theoretical performance guarantees.
	\subsubsection{The impact of background channel} To further examine the impact of the background channel, Fig.~\ref{fig:decay_h0} shows the average received SNR when the background channel is attenuated by a factor of $10^{-\alpha}$, where $\alpha > 0$ denotes a decay factor. When the background channel is strong ($\alpha \leq 1$), most algorithms still perform well. However, as $\alpha$ increases, i.e., as the background channel deteriorates, the performance gap between \textsc{CSM}, \textsc{RFocus}, and the optimal scheme becomes increasingly evident. Note that large $\alpha$ corresponds to the regime $h_0\approx 0$, thereby directly benchmarking the case when the background channel is weak. This indicates that both \textsc{RFocus} and \textsc{CSM} fail when the background channels are weak. This is because with a strong direct path, sensitivity to noise is reduced and voting-based mechanisms remain effective. In contrast, when the background channel weakens, random phase shifts may result in very low received SNR, making the problem substantially more challenging.

	\begin{figure}[t]
		\centering
		\includegraphics[width=0.5\textwidth]{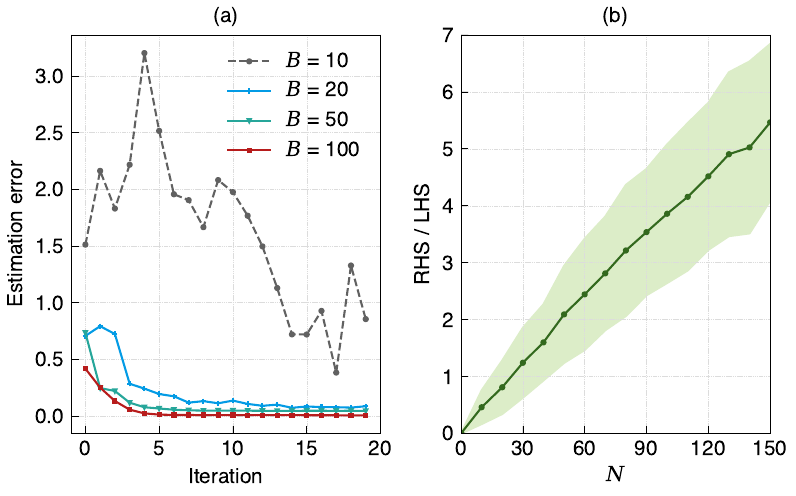}
		\caption{(a): The estimation error, defined as $\| \Mest - \Mtrue \|_2 + \| \west - \wtrue \|_2 $, over iterations in a LOS condition with different values of batch sizes: $B \in \{10, 20, 50, 100\}$. In this simulation, $N=5$; (b): The term \textsc{RHS}/\textsc{LHS} of Eq. \eqref{eq: Incoherence assumption} with different values of $N$. In this simulation, $B = 10 N$ and $N_s = 10 B$.}
		\label{fig:convergence_verify_assumption}
	\end{figure}
	\subsubsection{Performance benchmark in NLOS scenarios} Fig.~\ref{fig:NLOS_N20} and Fig.~\ref{fig:NLOS_N100} illustrate the CDF of the received SNR in the NLOS scenario with $N=20$ and $N=100$ reflective elements, respectively. In both cases, \textsc{CSM} and \textsc{RFocus} perform the worst, with results even lower than \textsc{RMS}. This observation is consistent with the previous result, and the findings in~\cite{10694808}, which proved that \textsc{CSM} and \textsc{RFocus} \textit{fail in environments with weak background channels} such as NLOS. For a small RIS (e.g., $N=20$), \textsc{GCSM} does not clearly outperform \textsc{RMS}. To be specific, it performs slightly better than \textsc{RMS} in the high-SNR regime, but worse in the low-SNR regime. This is because \textsc{GCSM} relies on constructing a virtual direct path by utilizing a partition of the RIS elements. While this approach can be effective when $N$ is sufficiently large, it is less effective when $N$ is small, as the resulting virtual direct path remains weak. In the case of a larger RIS ($N=100$), \textsc{GCSM} outperforms \textsc{RMS} more clearly. In both scenarios, \textsc{BORN} once again demonstrates its effectiveness by achieving nearly identical performance to the optimal benchmark, with approximately $7.0$~dB and $22.0$~dB at the $20$th percentile for $N=20$ and $N=100$, respectively. A clearer comparison among \textsc{BORN}, \textsc{GCSM}, and the optimal scheme under NLOS conditions is provided in Fig.~\ref{fig:NLOS_N150} where $N=150$. As observed, at the $40$th percentile, \textsc{BORN} attains near-optimal performance, which is nearly $1$~dB gain over \textsc{GCSM}.
	\subsubsection{The impact of noisy measurements and quantization} Now, we consider imperfect SNR measurements during the sensing stage. The ideal SNR is first perturbed by additive Gaussian noise with standard deviation $\sigma_{\mathrm{dB}}$, and then uniformly quantized with step size $\Delta$, both in dB. The resulting corrupted measurements are used in Stage~1 to estimate the quadratic model, while the final received SNR is still evaluated using the true channel. Fig.~\ref{fig:noisy_quantized_effect} shows the CDF of the received SNR for $\sigma_{\mathrm{dB}}=2$ dB with $\Delta=5$ dB and $\Delta=12$ dB. In both cases, \textsc{BORN} remains close to the optimal benchmark. However, compared with the ideal-measurement case, the gap between \textsc{BORN} and the optimal benchmark becomes larger as the quantization step increases from $\Delta=0$ to $\Delta=5$ dB and then to $\Delta=12$ dB. This is expected, since noisier and more coarsely quantized measurements reduce the accuracy of the QM estimation in Stage~1. Moreover, \textsc{BORN} also consistently outperforms all other baselines under both settings.
	\begin{figure}[t]
		\centering
		\includegraphics[width=0.44\textwidth]{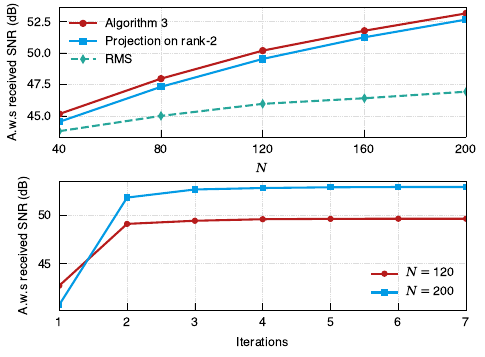}
		\caption{\textit{Top:} Average weighted sum (a.w.s) of the received SNR achieved by Algorithm~\ref{alg:alt_ascent}, rank-2 projection, and \textsc{RMS}. \textit{Bottom:} Evolution of the a.w.s received SNR over iterations for Algorithm~\ref{alg:alt_ascent}. In these simulations, $K_u = 4$.}
		\label{fig:multiuser}
	\end{figure}
	\subsubsection{Convergence analysis of Algorithm \ref{alg: gFM}} Fig.~\ref{fig:convergence_verify_assumption}(a) exhibits the estimation error across iterations in the sensing stage with different batch sizes for $N=5$ in LOS scenarios. In practice, $N$ is typically large; however, in this simulation we intentionally examine the convergence behavior for small $N$, as a low $N$ may violate the assumptions discussed in Assumption~\ref{Assp: Incoherence}. The results indicate that convergence can still be guaranteed if the batch size is chosen appropriately. As expected, increasing the batch size leads to improved convergence. When the batch size is small (e.g., $B=10$), convergence is not guaranteed, as this setting violates the condition in Theorem~\ref{theorem: main estimation error}. Intuitively, with such a small batch, the empirical quantities computed from the mini-batch have a large variance and may deviate substantially from their expectations. Since the update of Algorithm~\ref{alg: gFM} is essentially a stochastic process driven by these noisy estimates, the perturbations can be amplified across iterations rather than contracted, causing the iterates to drift away from $(\bm M^\star,\bm w^\star)$ and leading to an increasing estimation error. By contrast, when $B=100$, convergence is achieved within only 5 iterations, which is faster than the cases with $B=20$ or $B=50$.
	\subsubsection{Verification of Assumption~\ref{Assp: Incoherence}}
	Because the convergence guarantee of \textsc{BORN} is established under Assumption~\ref{Assp: Incoherence}, we empirically evaluate the ratio between the \textsc{RHS} and left-hand side (\textsc{LHS}) of~\eqref{eq: Incoherence assumption}. Fig.~\ref{fig:convergence_verify_assumption}(b) shows the sample mean of \textsc{RHS}/\textsc{LHS} versus $N$ together with a $\pm$1 standard-deviation band across trials. The ratio increases approximately linearly with $N$ and exceeds $1$ on average from $N= 30$. More importantly, the lower edge of the $\pm$1\, standard deviation band crosses $1$ around $N= 60$, after which the assumption holds with a comfortable margin (e.g., at $N= 140$ the mean is larger than $5$ and the lower band is larger than $3.5$). These indicate that the incoherence condition is reliably satisfied for moderate array sizes. We also note that \textsc{BORN} exhibits convergence even at smaller $N$ (as shown in Fig.~\ref{fig:convergence_verify_assumption}(a)), highlighting that the assumption is sufficient but not necessary in practice.
	\subsubsection{Benchmarking and convergence analysis for multi-user setting}
	We next evaluate the multi-user setting. Fig.~\ref{fig:multiuser} compares the average weighted sum (a.w.s) of the received SNR achieved by different methods and illustrates the convergence behavior of Algorithm~\ref{alg:alt_ascent}. In this experiment, we consider $K_u=4$ RXs, all located at the same distance from the RIS. Owing to this symmetric setup, all user weights are set to one. The top panel of Fig.~\ref{fig:multiuser} shows that Algorithm~\ref{alg:alt_ascent} consistently achieves the highest a.w.s of SNR over the entire range of $N$. The rank-2 projection method also provides a strong performance baseline, but it remains slightly inferior to Algorithm~\ref{alg:alt_ascent}. This is because, with $K_u=4$, $\bar{\bm M}$ can have rank up to $2K_u=8$. Therefore, restricting the design to a rank-2 approximation is generally insufficient to capture all dominant user directions. As a result, truncating $\bar{\bm M}$ to its two largest eigenmodes inevitably discards useful signal components associated with the remaining users, which leads to a loss in performance. The \textsc{RMS} method gives the lowest performance in all cases. The bottom panel of Fig.~\ref{fig:multiuser} demonstrates that Algorithm~\ref{alg:alt_ascent} converges rapidly. For both $N=120$ and $N=200$, most of the performance improvement is obtained within the first few iterations. This behavior confirms the efficiency of Algorithm \ref{alg:alt_ascent} in the multi-user setting.
	\subsubsection{Computational time}
	\begin{figure}[t]
		\centering
		\includegraphics[width=0.44\textwidth]{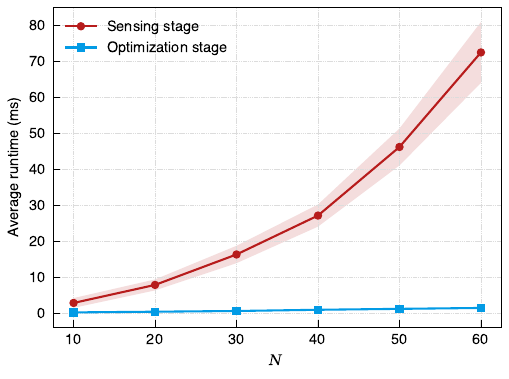}
		\caption{Average computational time of the sensing and optimization stages versus $N$. In this simulation, $B=8N$ and $N_s=10B$.}
		\label{fig:runtime}
	\end{figure}
	Fig.~\ref{fig:runtime} shows the average computational time\footnote{Note that the benchmark is obtained on a commodity laptop without hardware-specific optimization; hence, the reported values should be viewed as indicative reference runtimes rather than absolute deployment latency limits.} of the sensing and optimization stages with various values of $N$. Here, $B=8N$ and $N_s=10B$. Several observations can be made. First, the sensing stage is the dominant computational component, and its runtime increases noticeably with $N$, which is consistent with the higher complexity associated with learning the QM. By contrast, the optimization stage remains very lightweight across the entire tested range and only increases mildly with $N$. Even for $N=60$, the optimization stage stays at around the millisecond level, whereas the sensing stage requires several tens of milliseconds. Second, the gap between the two stages becomes larger as $N$ increases, which confirms that the proposed low-complexity optimization is not the computational bottleneck in practice. Overall, these runtime measurements complement the theoretical complexity analysis and support the practical feasibility of the proposed framework.
	\section{Conclusion} \label{sec:conclusion}
	In this paper, we introduce \textsc{BORN}, a blind beamforming algorithm for binary RIS that relies on RSS measurements. From a theoretical perspective, we derive sample complexity and suboptimality guarantees, showing that \textsc{BORN} attains near-optimal solutions with high probability. From an empirical perspective, we conduct field tests in a realistic manufacturing environment and simulations, which confirm the efficacy of \textsc{BORN}. The results highlight its robustness across both LOS and NLOS scenarios, outperforming existing blind beamforming methods. We also develop a learning framework for QM under Rademacher features and prove convergence.
	
	Future work includes extending \textsc{BORN} to RISs with multi-level discrete or continuous phase shifts, where the sensing-optimization framework remains applicable but the current analysis must be substantially redeveloped. Another important direction is joint transceiver-beamformer and RIS optimization in multi  antenna systems, as well as extensions to more rapidly time-varying channels.
	
	\vspace{-0.3cm}
	\section*{Acknowledgment}
	The authors thank Mr.~Alex Schofield and Mr.~Wass Waseem for their valuable support during the field tests.
	\vspace{-0.2cm}
	\appendix
	\textit{Additional Notations:} We introduce global notations, which are used consistently throughout the appendix.
	Given a constant $\delta$ and an incoherence constant $\bar \mu_0$, we define a constant
	\begin{equation}
		C(\delta, \bar{\mu}_0) \triangleq \RoundBr{\frac{\bar \mu_0 r}{N} + 2\delta}.
	\end{equation}
	
	Let \((e_1, \dots, e_d)\) denote the canonical basis of the vector space \(\mathbb{R}^d\). For a symmetric matrix $\bm M$, let $\lambda_k(\bm M)$ be the $k$-th largest eigenvalue of $\bm M$.
	Let \(\bm U \in \mathbb{R}^{N\times r}\) with orthonormal columns, e.g., \(\bm U^\top \bm U = \bm I_r\), an orthonormal complement of \(\bm U\) is any matrix
	\(\bm U_{\bot} \in \mathbb{R}^{N\times (N-r)}\) whose columns form an orthonormal
	basis of the null space of $\bm U^\top$; equivalently,
	\begin{equation}
		\begin{aligned}
			&\bm U^\top \bm U_{\bot} = \bm 0_{r \times (N-r)}, \quad \quad \bm U_{\bot}^\top \bm U_{\bot} = \bm I_{N-r},\; \\
			&[\bm U ,  \bm U_{\bot}] \text{ is orthonormal}.
		\end{aligned}
	\end{equation}
	For 2 matrices $\bm U$ and $\bm V$, let $\theta(\bm U, \bm V)$ be the largest canonical angle between the column spaces of $\bm U$ and $\bm V$, that is, 
	\begin{align}
		\sin \theta(\bm U, \bm V) &= \|\bm U_{\bot}^\top \bm V\|_2,\quad \nonumber \\
		\cos \theta(\bm U, \bm V) &= \sigma_r(\bm U^\top \bm V), \\
		\tan \theta(\bm U, \bm V) &= \|\bm U_{ \bot}^\top \bm V(\bm U^\top \bm V)^{-1}\|_2. \nonumber
	\end{align}
	Denote the eigen-decomposition of $\Mtrue$ as $\Mtrue = \bm U_\star \bm \Lambda_\star \bm U_\star^\top$ where $\bm \Lambda_\star = \Diag{\lambda_1(\Mtrue),\dots,\lambda_r(\Mtrue)}$
	then for the sequence of estimation matrix $\bm U_t$ in Algorithm \ref{alg: gFM}, we denote $\theta_t \triangleq \theta(\bm U_\star,\bm U_t)$.
	By convention, we write \(\bm{M} = \widehat{\bm{M}} + O(\epsilon)\) to mean that
	$
	\|\bm{M} - \widehat{\bm{M}}\|_2 = O(\epsilon).
	$
	
	Before starting the main proofs, note that we sometimes introduce additional notation that is defined explicitly within individual proofs. Inspired by programming conventions, these notations are local, meaning they are only valid within the body of the corresponding proof. 
	
	
	\subsection{Proof of Proposition~\ref{prop: Projection to PSD matrices}} \label{proof: Projection to PSD matrices}
	Because $\bm M$ is symmetric, it admits an eigen-decomposition $\bm M=\bm Q \bm D \bm Q^\top$ with $\bm Q$ orthogonal and $\bm D$ diagonal. 
	Let $\bm M'\in\mathcal S_+^N$ with $\Rank{\bm M'}\leq r$ be any feasible point. Define
	$
	\bm Y \triangleq \bm Q\,\bm M'\,\bm Q^\top,
	$
	then $\bm Y$ is symmetric, PSD, and $\Rank{\bm Y}\le r$. Using orthogonal invariance of the Frobenius norm, we have
	\begin{align*}
		\|\bm M' - \bm M\|_F
		= \|\bm Q \bm M' \bm Q^\top - \bm Q \bm M \bm Q^\top\|_F
		= \|\bm Y - \bm D\|_F.
	\end{align*}
	\[
	\text{Hence, }\min_{\substack{\bm M' \in \mathcal S_+^N\\ \Rank{\bm M'}\le r}}
	\|\bm M' - \bm M\|_F
	\;\;=
	\min_{\substack{\bm Y \succeq 0\\ \Rank{\bm Y}\leq r}}
	\|\bm D - \bm Y\|_F.
	\]
	Write $\bm Y=(y_{i,j})_{i,j=1}^N$. Since $\bm D$ is diagonal,
	\begin{align} \label{eq: projecting wrt F norm}
		\|\bm D - \bm Y\|_F^2
		= \sum_{i=1}^N (\lambda_i - y_{ii})^2 + \sum_{i\neq j} y_{ij}^2.
	\end{align}
	Any off-diagonal $y_{ij}$ (for $i\neq j$) only increases the objective by $y_{ij}^2\ge0$ while not reducing any diagonal discrepancy since $\bm D$ has zeros off-diagonal. Therefore, at optimality, $y_{ij}=0$ for $i\neq j$, and the minimizer is diagonal:
	$
	\bm Y = \Diag{y_1,\ldots,y_N}.
	$
	The optimization problem in \eqref{eq: projecting wrt F norm} is equivalent to
	\[
	\min_{\substack{y_i\ge 0\\ \text{at most $r$ nonzeros}}}
	\sum_{i=1}^N (\lambda_i - y_i)^2.
	\]
	This yields $\bm Y^\star = \Diag{\lambda_1^+,\ldots,\lambda_r^+,\bm 0_{N-r}} = \bm D^+.$
	Finally, transforming back gives
	$
	\bm M^+ = \bm Q^\top \bm Y^\star \bm Q = \bm Q^\top \bm D^+ \bm Q,
	$
	which is symmetric, PSD, rank $\le r$, and achieves the minimum Frobenius distance to $\bm M$. This proves the claim.
	\qed
	
	\subsection{Proof of Theorem \ref{theorem: main estimation error}} \label{proof of theorem main estimation error}
	Before presenting the proof of Theorem \ref{theorem: main estimation error}, we first introduce important supporting lemmas.
	\begin{lemma} \label{lem: H and estimation error}
		Let $\bm H_1^{(t)}, h_2^{(t)},  \bm h_3^{(t)}$ be defined in Algorithm \ref{alg: gFM}. Let $\epsilon_t \triangleq \|\bm M_t - \Mtrue \|_2 + \|\bm w_t - \bm w_\star \|_2 $. Then with probability at least $1-\eta$, provided $B = Cr^3N\log_2(\eta^{-1})/\delta^2$ for some universal constant $C>0$, 
		\begin{align} \label{eq:H123}
			&\!\!\Norm{\bm H_1^{(t)} \!- \! (\Mtrue\! - \! \bm M_t) \!+ \! \Tr{\Mtrue \!-\! \bm M_t}  
				\!+ \!\Diag{\Mtrue \!- \!\bm M_t}}_2 \!\leq \delta\epsilon_t, \nonumber\\
			&|h_2^{(t)} - \Tr{\Mtrue - \bm M_t}| \leq \delta\epsilon_t,  \\
			&\Norm{\bm h_3^{(t)} - \bm w_\star + \bm w_t}_2 \leq \delta \epsilon_t. \nonumber
		\end{align}
	\end{lemma}
	\begin{proof}
		The proof follows by applying Lemma 6 of \cite{lin2017second} and noting that Rademacher is a sub-Gaussian distribution.
	\end{proof}
	The convergence of Algorithm \ref{alg: gFM} is characterized by the decreasing error between the column subspaces of $\Mtrue$ and $\bm M_t$.
	We first state the error due to the unrecoverable diagonal entries of $\Mtrue$.
	\begin{lemma} \label{lem:estimation of Mtrue}
		Suppose Assumption \ref{Assp: Incoherence} holds, then
		\begin{equation}\label{eq: bound diag error}
			\|\Diag{\widetilde{\bm M}_t}\|_2 \leq \frac{\bar\mu_0\, r}{N}\|\widetilde{\bm M}_t\|_2,
		\end{equation}
		where $\widetilde{\bm M}_t \triangleq \Mtrue - \bm M_t$. Moreover, given the batch size $B = Cr^3N\log_2(\eta^{-1})/\delta^2$ for some universal constant $C>0$, with probability at least $1-\eta$, 
		\begin{equation} \label{eq: estimate Mtrue by H}
			\| \bm M_t + \! \bm H_1^{(t)} \!  \!  + h_2^{(t)}\bm I -  \! \Mtrue \|_2 
			\leq C(\delta, \bar{\mu}_0) \epsilon_t.
		\end{equation}
	\end{lemma}
	\begin{proof}
		Let $\widetilde{\bm M}_t \triangleq \Mtrue - \bm M_t$, and rewrite $\widetilde{\bm M}_t = \widetilde{\bm U}_t \widetilde{\bm \Lambda}_t  \widetilde{\bm U}_t^\top$. Under Assumption \ref{Assp: Incoherence}, $\bm{\tilde M}_t$ is $\bar \mu_0$-incoherent. Hence,
		\begin{align*} 
			&\|\Diag{\widetilde{\bm M}_t}\|_2
			= \max_i \big|\bm e_i^\top \widetilde{\bm M}_t \bm e_i\big| \nonumber\\
			&= \max_i \big|(\widetilde{\bm U}_t^\top \bm e_i)^\top \widetilde{\bm\Lambda}_t (\widetilde{\bm U}_t^\top \bm e_i)\big| \overset{(i)}{\le}\; \|\widetilde{\bm\Lambda}_t\|_2 \;\max_i \|\widetilde{\bm U}_t^\top \bm e_i\|_2^2 \\&\overset{(ii)}{=}\; \|\widetilde{\bm M}_t\|_2 \;\max_i \|\widetilde{\bm U}_{t,i:}\|_2^2 \overset{(iii)}{\le}\; \|\widetilde{\bm M}_t\|_2 \;\frac{\bar\mu_0\, r}{N}, \nonumber
		\end{align*}
		where $(i)$ is due to the Rayleigh–type bound property for symmetric $\bm A$, $|\bm x^\top \bm A \bm x|\le \|\bm A\|_2\,\|\bm x\|_2^2$; and $(ii)$ holds because $\|\widetilde{\bm\Lambda}_t\|_2=\|\tilde{\bm M}_t\|_2$ and $\|\tilde{\bm U}_t^\top \bm e_i\|_2=\|\tilde{{\bm U}}_{t,i:}\|_2$. Finally, $(iii)$ is due to Assumption \ref{Assp: Incoherence}.

		The second part, \eqref{eq: estimate Mtrue by H}, can be proved by using a triangle inequality combined with \eqref{eq: bound diag error} and Lemma \ref{lem: H and estimation error} as follows
		\begin{align}
			&\! \!  \!  \! 	\| \bm M_t + \! \bm H_1^{(t)} \!  \!  + h_2^{(t)}\bm I -  \! \Mtrue \|_2 
			\leq \|\Diag{\Mtrue - \bm M_t}\|_2 + 2\delta \epsilon_t  \nonumber\\
			&\leq \frac{\bar \mu_0 r}{N} \epsilon_t + 2\delta \epsilon_t = C(\delta, \bar{\mu}_0) \epsilon_t.
		\end{align}
		The proof is now completed.
	\end{proof}
	\color{black}
	
	\begin{lemma}[\textbf{Decreasing estimation error}] \label{lemma: Decreasing Error with Coherence Assumption} 
		Let \\
		$\alpha_t \triangleq \tan \theta_t$, $\beta_t \triangleq \|\Mtrue - \bm M_t \|_2$, $\gamma_t \triangleq \|\wtrue - \bm w_t \|_2$, and $\epsilon_t \triangleq \beta_t + \gamma_t$.
		Suppose Assumption \ref{Assp: Incoherence} holds. Moreover, suppose $\alpha_t \leq 2$, and $C(\delta, \bar{\mu}_0) \epsilon_t \leq \tfrac{\sigma_r(\Mtrue)}{2\sqrt{5}}$. 
		Then,
		\begin{equation}
			\begin{aligned}
				\alpha_{t+1} &\leq 2\sqrt{5} \;(\sigma_r(\Mtrue ))^{-1} C(\delta, \bar{\mu}_0) \epsilon_t,\\
				\beta_{t+1} & \leq  \alpha_{t+1} \|\Mtrue \|_2 +  C(\delta, \bar{\mu}_0) \epsilon_t, \\
				\gamma_{t+1} &\leq  C(\delta, \bar{\mu}_0) \epsilon_t.
			\end{aligned}
		\end{equation}
	\end{lemma}
	\begin{proof}
		\textit{First}, we bound $\alpha_{t+1}$.
		Consider 
		\begin{equation}
			\begin{aligned} \label{eq: bound U* Ut}
				&\!\!\!\!\! \Norm{\bm U_{\star, \bot}^\top \bm U_{t+1}}_2 
				= \Norm{\bm U_{\star, \bot}^\top \; (\bm M_t + \! \bm H_1^{(t)} \!  \!  + h_2^{(t)}\bm I) \; \bm U_t}_2 \\
				&\overset{(i)}{\leq} \Norm{\bm U_{\star, \bot}^\top \RoundBr{\Mtrue+ C(\delta, \bar{\mu}_0) \epsilon_t} \bm U_t}_2 \\
				&\overset{(ii)}{\leq}  \Norm{\bm U_{\star, \bot}^\top \Mtrue \bm U_t}_2 + C(\delta, \bar{\mu}_0) \epsilon_t 
				=C(\delta, \bar{\mu}_0) \epsilon_t.
			\end{aligned}
		\end{equation}
		In \eqref{eq: bound U* Ut}, $(i)$ is due to Lemma \ref{lem:estimation of Mtrue}; and $(ii)$ is because $\bm U_{\star, \bot}$ and $\bm U_t$ are orthonormal and their operator norm is $1$. Next, 
		\begin{equation}
			\begin{aligned} \label{eq: bound sigma_r of U* U_t}
				\!\!\!\! \sigma_r\!\RoundBr{\bm U_\star ^\top \bm U_{t+1}} &= \sigma_r\RoundBr{\bm U_{\star}^\top (\bm M_t + \! \bm H_1^{(t)} \!  \!  + h_2^{(t)}\bm I) \bm U_t} \\ 
				&\overset{(i)}{\geq} \sigma_r\!\RoundBr{ \bm U_{\star}^\top \RoundBr{\Mtrue - C(\delta, \bar{\mu}_0) \epsilon_t}\bm U_t} \\
				&\geq \sigma_r( \bm U_{\star}^\top \Mtrue \bm U_t) -  C(\delta, \bar{\mu}_0) \epsilon_t \\
				&\geq \sigma_r(\Mtrue ) \sigma_r(\bm U_{\star}^\top \bm U_t) - C(\delta, \bar{\mu}_0) \epsilon_t \\
				&= \sigma_r(\Mtrue ) \cos \theta_t - C(\delta, \bar{\mu}_0) \epsilon_t. 
			\end{aligned}
		\end{equation}
		In \eqref{eq: bound sigma_r of U* U_t}, $(i)$ is due to Lemma \ref{lem:estimation of Mtrue} and the fact that $\sigma_r(A-B) \geq \sigma_r(A) - \sigma_1(B)$ for any two matrices of same size $A, \;B$.
		Therefore, we can bound $\alpha_{t+1}$ as
		\begin{equation} \label{eq: bounding tan theta 1}
			\begin{aligned}
				\alpha_{t+1} &= \|\bm U_{ \star, \bot}^\top \bm U_{t+1}(\bm U_\star ^\top \bm U_{t+1})^{-1}\|_2 \overset{(i)}{\leq} \frac{\|\bm U_{ \star, \bot}^\top \bm U_{t+1}\|_2}{\sigma_r\RoundBr{\bm U_\star ^\top \bm U_{t+1}}} \\
				&\overset{(ii)}{\leq} \frac{C(\delta, \bar{\mu}_0) \epsilon_t}{\sigma_r(\Mtrue ) \cos \theta_t - C(\delta, \bar{\mu}_0) \epsilon_t}, \\
			\end{aligned}
		\end{equation}
		where $(i)$ is due to submultiplicativity and $\|B^{-1}\|_2=1/\sigma_r(B)$, i.e., $\|AB^{-1}\|_2 \le \|A\|_2/\sigma_r(B)$, and $(ii)$ is due to \eqref{eq: bound U* Ut} and \eqref{eq: bound sigma_r of U* U_t}.
		
		Since \(\alpha_t = \tan \theta_t \le 2\) and \(\theta_t \in [0, \pi/2]\), we have \(\cos \theta_t \ge 1/\sqrt{5}\).
		Moreover, using the assumption that
		$
		C(\delta, \bar{\mu}_0) \epsilon_t \leq \frac{\sigma_r(\Mtrue )}{2\sqrt{5}} $ and substituting into \eqref{eq: bounding tan theta 1}, we obtain
		\begin{equation} 
			\begin{aligned}
				\alpha_{t+1} \leq 2\sqrt{5} \; (\sigma_r(\Mtrue ))^{-1} C(\delta, \bar{\mu}_0) \epsilon_t.
			\end{aligned}
		\end{equation}
		
		\textit{Next}, we bound $\beta_t$.
		First, according to definition of $\bm M_{t+1}$ and $\bm V_{t+1}$ in Algorithm \ref{alg: gFM}, we can show that
		\begin{align*}
			\Norm{\Mtrue \!-\! \bm M_{t+1}}_2 \leq  \Norm{\Mtrue \!-\!\bm U_{t+1} \bm U_{t+1}^\top \RoundBr{\!\bm H_1^{(t)} \!-\! h_2^{(t)}\bm I \!+\! \bm M_t}\!}_2.
		\end{align*}
		
		Now, we can bound $\beta_{t+1}$ as follows.
		\begin{align*}
			&\beta_{t+1} = \Norm{\Mtrue - \bm M_{t+1}}_2 \\
			& = \Norm{\Mtrue - \bm U_{t+1} \bm U_{t+1}^\top \RoundBr{\bm H_1^{(t)} - h_2^{(t)}\bm I + \bm M_t}}_2 \nonumber \\
			& \overset{(i)}{\leq} \Norm{\Mtrue - \bm U_{t+1} \bm U_{t+1}^\top \RoundBr{\Mtrue + C(\delta, \bar{\mu}_0) \epsilon_t }}_2 \nonumber\\
			& \overset{(ii)}{\leq} \Norm{(\bm I - \bm U_{t+1} \bm U_{t+1}^\top) \; \Mtrue}_2     +  C(\delta, \bar{\mu}_0) \epsilon_t    \nonumber\\
			&\leq \alpha_{t+1} \|\Mtrue \|_2 +  C(\delta, \bar{\mu}_0) \epsilon_t ,
		\end{align*}
		where $(i)$ is due to Lemma \ref{lem:estimation of Mtrue} while $(ii)$ holds since $\bm U_{t+1}$ has orthonormal column and its operator norm is $1$. 

		\textit{Finally}, the bound of $\gamma_{t+1}$ can be obtained by following similar transformation. The proof is now completed.
	\end{proof}
	Lemma~\ref{lemma: Decreasing Error with Coherence Assumption} indicates that the estimation error decreases over iterations if $2$ following conditions hold: (a) $C(\delta, \bar{\mu}_0)\epsilon_t \le \tfrac{\sigma_r(\Mtrue)}{2\sqrt{5}}$, and (b) $\alpha_t \le 2$. We will show below that both conditions are satisfied by choosing a sufficiently small $\delta$.
	In particular, the choice of \(\delta\) specified in \eqref{eq: delta to gurantee convergence 2} guarantees the condition (a), while the condition (b) is satisfied in all iterations via \eqref{eq: delta to gurantee convergence 3}. The only remaining task is to verify the initialization of \(\alpha_0\). Specifically, we bound \(\alpha_0\) as follows.
	
	\begin{lemma}[\textbf{Error at initialisation}] \label{lem: init error}
		Suppose Assumption \ref{Assp: Incoherence} holds.
		Choose $\delta$ be sufficiently small so that
		\begin{equation} \label{eq: choice of delta for small init error}
			\delta \;\le\; \frac{\sigma_r(\Mtrue)}{16\bigl(\|\Mtrue\|_2 + \|\wtrue\|_2\bigr)}.
		\end{equation}
		Under these conditions, the initial error is bounded as \(\alpha_0 \le 2\).
	\end{lemma}
	\begin{proof}
		
		
		We begin by bounding the error at initialization as
		\begin{align*}
			\epsilon_0 &= \Norm{\! \bm H_1^{(0)} \! \! + h_2^{(0)}\bm I - \! \Mtrue}_2 + \|\bm h_3^{(0)}-\wtrue \|_2 \\
			&\! \!\leq  \!\Norm{\! \bm H_1^{(0)} \! +\! h_2^{(0)}\bm I \!- \! \Mtrue\! + \!\Diag{\Mtrue}  \!- \!\Diag{\Mtrue}\!}_2 \nonumber
			\!+ \\
			& \quad \!\|\bm h_3^{(0)}-\wtrue \|_2 \\
			&  \!\!\overset{(i)}{\leq} \delta\|\Mtrue \|_2 + \frac{\bar \mu_0 r}{N}\|\Mtrue \|_2 + \delta\epsilon\|\wtrue \|_2 \nonumber \\
			& \!\!\leq C(\delta, \bar{\mu}_0)(\|\Mtrue \|_2 + \|\wtrue \|_2), \nonumber
		\end{align*}
		where the inequality $(i)$ is due to Lemma \ref{lem: H and estimation error} and \eqref{eq: bound diag error}.
		
		According to Lemma 12 introduced in \cite{Lin2016gFM}, at initialization, we can choose a sufficiently small $\delta$  so that 
		\begin{align}
			&\epsilon_0 &&\leq C(\delta, \bar{\mu}_0)(\|\Mtrue \|_2 + \|\wtrue \|_2) \leq \frac{\sigma_r(\Mtrue)}{4}, \nonumber\\
			\Longrightarrow \; &  \sin \theta_0 &&\leq 2\frac{C(\delta, \bar{\mu}_0)(\|\Mtrue \|_2 + \|\wtrue \|_2)}{\sigma_r(\Mtrue)}.
		\end{align}
		Therefore, $\alpha_0 \leq 2$ provided that $\sin \theta_0 \leq 1/2$, that is, 
		\begin{equation} \label{eq: condition of delta for init error 1}
			C(\delta, \bar{\mu}_0) \leq \frac{\sigma_r(\Mtrue)}{4(\|\Mtrue \|_2 + \|\wtrue \|_2)}.
		\end{equation}
		Now, by Assumption \ref{Assp: Incoherence} with constant in $O$-notation is $8$, that is,
		$
		\frac{\bar \mu_0 r}{N}\leq \frac{\sigma_r(\Mtrue)}{8(\|\Mtrue \|_2 + \|\wtrue \|_2)}.
		$
		To guarantee \eqref{eq: condition of delta for init error 1}, we need
		$
		\delta \leq \frac{\sigma_r(\Mtrue)}{16(\|\Mtrue \|_2 + \|\wtrue \|_2)}.
		$
		In summary, we have $\alpha_0 \leq 2$ provided that $\delta \leq \frac{\sigma_r(\Mtrue)}{16(\|\Mtrue \|_2 + \|\wtrue \|_2)}$, which concludes the proof.
		\color{black}
	\end{proof}
	
	We now proceed with the analysis of Algorithm \ref{alg: gFM}.
	\begin{proof}[\textbf{Proof of Theorem \ref{theorem: main estimation error}}]
		\textbf{Step 1.} We first prove the error convergence of after each iteration.
		Suppose that at step $t$ , $C(\delta, \bar{\mu}_0)\epsilon_t \leq \tfrac{\sigma_r(\Mtrue)}{4\sqrt{5}}$ and $\alpha_t\leq 2$, from \emph{Lemma \ref{lemma: Decreasing Error with Coherence Assumption}}, we have 
		\begin{align}
			\epsilon_{t+1} &= \beta_{t+1} + \gamma_{t+1} \nonumber
			\leq \gamma_{t+1} + \alpha_{t+1} \|\Mtrue \|_2 +  C(\delta, \bar{\mu}_0) \epsilon_t \\
			&\leq C(\delta, \bar{\mu}_0) \epsilon_t + 2\sqrt{5}\frac{\sigma_1(\Mtrue )}{\sigma_2(\Mtrue )}C(\delta, \bar{\mu}_0) \epsilon_t + C(\delta, \bar{\mu}_0) \epsilon_t  \nonumber\\
			&\leq \frac{7\sigma_1(\Mtrue )}{\sigma_2(\Mtrue )}C(\delta, \bar{\mu}_0)\epsilon_t. \nonumber
		\end{align}
		Therefore,
		\begin{align} \label{eq: closed form of error}
			\epsilon_{t} & \leq \RoundBr{\frac{7\sigma_1(\Mtrue )}{\sigma_2(\Mtrue )}C(\delta, \bar{\mu}_0)}^t (\|\Mtrue \|_2 + \| \wtrue\|_2), \nonumber\\
			\alpha_{t+1} &\overset{(i)}{\leq} 2\sqrt{5}\sigma_r^{-1}(\Mtrue )C(\delta, \bar{\mu}_0)  \\
			&\quad \times \RoundBr{\frac{7\sigma_1(\Mtrue )}{\sigma_2(\Mtrue )} 
				C(\delta, \bar{\mu}_0)}^t (\|\Mtrue \|_2 + \| \wtrue\|_2) \nonumber,
		\end{align}
		where $(i)$ is due to Lemma \ref{lemma: Decreasing Error with Coherence Assumption} and the bound on $\epsilon_t$.
		
		\textbf{Step 2.} Based on \eqref{eq: closed form of error}, we derive the condition for convergence.
		First, by the Assumption \ref{Assp: Incoherence} with constant in $O$-notation is $28$, that is,
		\begin{equation}\label{eq: explicit const in A1}
			\frac{\bar \mu_0 r}{N}\leq \frac{\sigma_r(\Mtrue)}{28(\|\Mtrue \|_2 + \|\wtrue \|_2)},
		\end{equation}
		to guarantee convergence, we need that
		\begin{equation} \label{eq: delta to gurantee convergence 1}
			\frac{7\sigma_1(\Mtrue )}{\sigma_2(\Mtrue )}C(\delta, \bar{\mu}_0) \leq 1 \Longleftarrow \delta \leq \frac{\sigma_2(\Mtrue)}{28\sigma_1(\Mtrue)}.
		\end{equation}
		
		Second, by closed form of $\alpha_{t+1}$ in \eqref{eq: closed form of error}, to ensure condition $\alpha_{t+1} \leq 2$ recursively, we need $\alpha_0 \leq 2$ which can be guaranteed in Lemma \ref{lem: init error} with the choice of $\delta$ as in \eqref{eq: choice of delta for small init error} and
		\begin{equation} \label{eq: delta to gurantee convergence 2}
			\begin{aligned}
				&2\sqrt{5}\frac{\|\Mtrue \|_2 + \|\wtrue\|_2}{\sigma_2(\Mtrue )}C(\delta, \bar{\mu}_0) \leq 2 \\
				\Longleftarrow \quad  &\delta \leq \frac{\sigma_2(\Mtrue)}{14 (\|\Mtrue \|_2 + \|\wtrue\|_2)}.
			\end{aligned}
		\end{equation}
		
		Third, we need to ensure the condition that $C(\delta, \bar{\mu}_0) \epsilon_t \leq \tfrac{\sigma_r(\Mtrue)}{2\sqrt{5}}$ used by Lemma \ref{lemma: Decreasing Error with Coherence Assumption}.
		Now, by Assumption \ref{Assp: Incoherence} with explicit constant as in \eqref{eq: explicit const in A1}, and note that $\epsilon_t \leq \epsilon_0$ due to \eqref{eq: closed form of error} and the choice of $\delta$ made in \eqref{eq: delta to gurantee convergence 1}, we have that
		\(   \frac{\bar{\mu}_0r}{N} \epsilon_t \leq\frac{\sigma_r(\Mtrue)}{4\sqrt{5}}.
		\)
		Therefore, to guarantee  $C(\delta, \bar{\mu}_0) \epsilon_t \leq \tfrac{\sigma_r(\Mtrue)}{2\sqrt{5}}$, we need
		\begin{equation} \label{eq: delta to gurantee convergence 3}
			\delta \leq \frac{\sigma_2(\Mtrue)}{4\sqrt{5}(\|\Mtrue \|_2 + \|\wtrue\|_2)}.
		\end{equation}
		
		\color{black}
		In summary, by \eqref{eq: choice of delta for small init error}, \eqref{eq: delta to gurantee convergence 1}, \eqref{eq: delta to gurantee convergence 2},  \eqref{eq: delta to gurantee convergence 3}, we need
		\begin{equation}
			\begin{aligned}
				\delta &\leq \min\Bigg\{\frac{\sigma_r(\Mtrue)}{16(\|\Mtrue \|_2 + \|\wtrue \|_2)}, \frac{\sigma_2(\Mtrue)}{28\sigma_1(\Mtrue)}, \\
				& \quad \quad \frac{\sigma_2(\Mtrue)}{14 (\|\Mtrue \|_2 + \|\wtrue\|_2)},
				\frac{\sigma_2(\Mtrue)}{4\sqrt{5}(\|\Mtrue \|_2 + \|\wtrue\|_2)} \Bigg\}; \\
				\Longleftarrow \delta &= \frac{\sigma_2(\Mtrue)}{28\sigma_1(\Mtrue) + 16\|\wtrue\|_2} .
			\end{aligned}
		\end{equation}
		With this choice of $\delta$, then for any $\varepsilon>0$ given the parameters stated in the theorem, we have that
		\begin{equation}
			\|\Mtrue - \bm M_T\|_2 + \| \wtrue - \bm w_T\|_2 \leq \varepsilon.
		\end{equation}
		
		\textbf{Step 3.} We proved the error due to projection of $\bm M_T$ to space of PSD matrices of rank at most $r$.
		We have that 
		\begin{equation}
			\|\Mest -\Mtrue \|_F \leq  \|\Mest - \bm M_T \|_F + \|\bm M_T - \Mtrue \|_F.
		\end{equation}
		Note that since $\Mtrue$ lies in the space of PSD matrices of rank at most $r$, $\|\Mest - \bm M_T \|_F \leq \|\Mtrue - \bm M_T \|_F$.
		Therefore,
		\begin{equation}
			\begin{aligned}
				\|\Mest -\Mtrue \|_2 &\leq \|\Mest -\Mtrue \|_F \leq  2\|\bm M_T - \Mtrue \|_F \\
				&\leq 6r \|\bm M_T - \Mtrue \|_2 \leq 6r\varepsilon,
			\end{aligned}
		\end{equation}
		since $\bm M_T$ and $\bm M_T - \Mtrue$ have rank at most $2r$ and $3r$, respectively.
	\end{proof}
	\vspace{-0.3cm}
	\subsection{Proof of Lemma \ref{lemma: estimation error to optimal gap}} \label{proof of lemma estimation error to optimal gap}
	For $\forall \bm{x} \in \{-1,1 \}^N$, we have
	\begin{equation} \label{eq: gap due to est error of M and w}
		\begin{aligned}
			y(\bm x) - \widehat{y}(\bm x) &= |\bm{x} ^\top  (\Mtrue - \Mest)  \bm{x} + \bm{x}^\top (\wtrue - \west)| \\
			& \leq |\bm{x} ^\top  (\Mtrue - \Mest)  \bm{x}| + |\bm{x}^\top (\wtrue - \west)| \\
			&\leq \| \Mtrue - \Mest\|_2 \| \bm{x}\|_2^2 + \|\wtrue - \west\|_2 \| \bm{x}\|_2 \\
			&\leq N\epsilon_M + \sqrt{N} \epsilon_w.
		\end{aligned}
	\end{equation}
	Therefore, we have the following inequalities
	\begin{equation}
		\begin{aligned}
			&\xtrue^\top \Mtrue  \xtrue + \xtrue^\top \wtrue \\
			&\leq \xtrue^\top \Mest \xtrue +  \xtrue^\top \west + N\epsilon_M + \sqrt{N} \epsilon_w\\
			&\leq \xest^\top \Mest \xest +  \xest^\top \west +N\epsilon_M + \sqrt{N} \epsilon_w\\
			&\leq \xest^\top \Mtrue \xest +  \xest^\top \wtrue +2N\epsilon_M + 2\sqrt{N} \epsilon_w, \\
		\end{aligned}
	\end{equation}
	where the first and third inequalities are due to \eqref{eq: gap due to est error of M and w}, and the second inequality holds thanks to Lemma \ref{lemma:suboptimality_gap_optimization}.
	\qed
	\subsection{Proof of Theorem \ref{theorem: main optimal gap}} \label{proof of theorem optimal gap}
	The proof of Theorem \ref{theorem: main optimal gap} is achieved directly by combining Theorem \ref{theorem: main estimation error} and Lemma \ref{lemma: estimation error to optimal gap}. By Theorem \ref{theorem: main estimation error}, we can bound the estimation error as
	\begin{equation}
		\begin{aligned}
			\|\Mest - \Mtrue\|_2 \leq \frac{6\varepsilon}{14N}, \quad \quad \|\west - \wtrue\|_2 \leq \frac{\varepsilon}{14Nr}.
		\end{aligned}
	\end{equation}
	Now, consider Lemma \ref{lemma: estimation error to optimal gap}, let $\epsilon_M = \tfrac{6\varepsilon}{14N}$ and $\epsilon_w = \tfrac{\varepsilon}{14Nr}$.
	By Lemma \ref{lemma: estimation error to optimal gap}, we conclude that
	\begin{equation}
		\begin{aligned}
			y(\bm x_\star)  - y(\widehat{\bm x}) & \leq 2N\epsilon_M + 2\sqrt{N} \epsilon_w \leq \varepsilon.
		\end{aligned}
	\end{equation}
	The proof is now completed. \qed
	
	\bibliographystyle{IEEEtran}
	\bibliography{bibfile_gen}
	
	\begin{IEEEbiography}
		[{\includegraphics[width=1in,height=1.25in,clip,keepaspectratio]{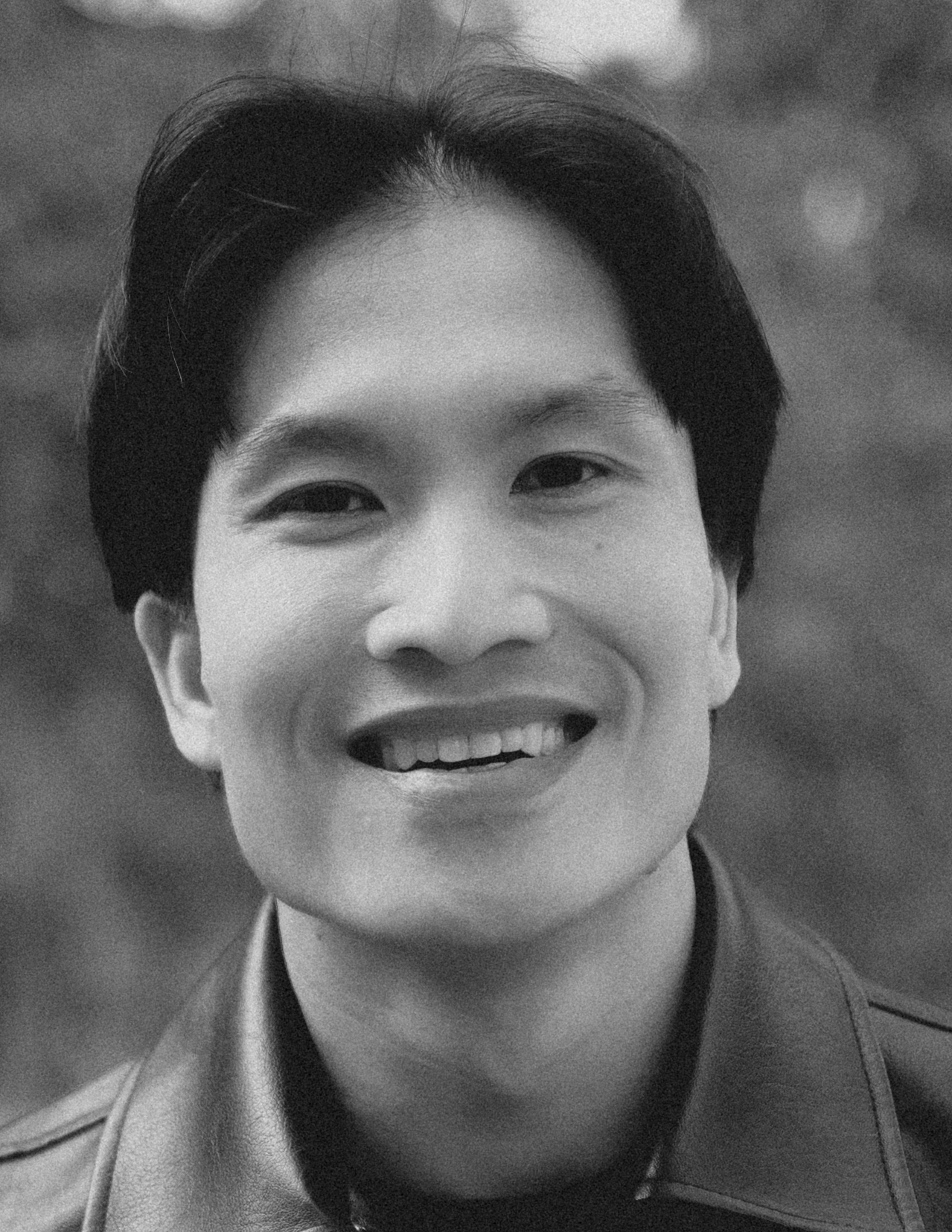}}]{Son Dinh-Van} (Member, IEEE)
		is an Assistant Professor with WMG, the University of Warwick, U.K. His research bridges communication theory and practical 5G/6G connectivity, with particular emphasis on signal processing and optimisation for multiple-antenna systems, reconfigurable intelligent surfaces, non-terrestrial networks, and AI-native radio access networks. He is also an Associate Fellow of the Higher Education Academy.
	\end{IEEEbiography}

	\begin{IEEEbiography}
		[{\includegraphics[width=1in,height=1.25in,clip,keepaspectratio]{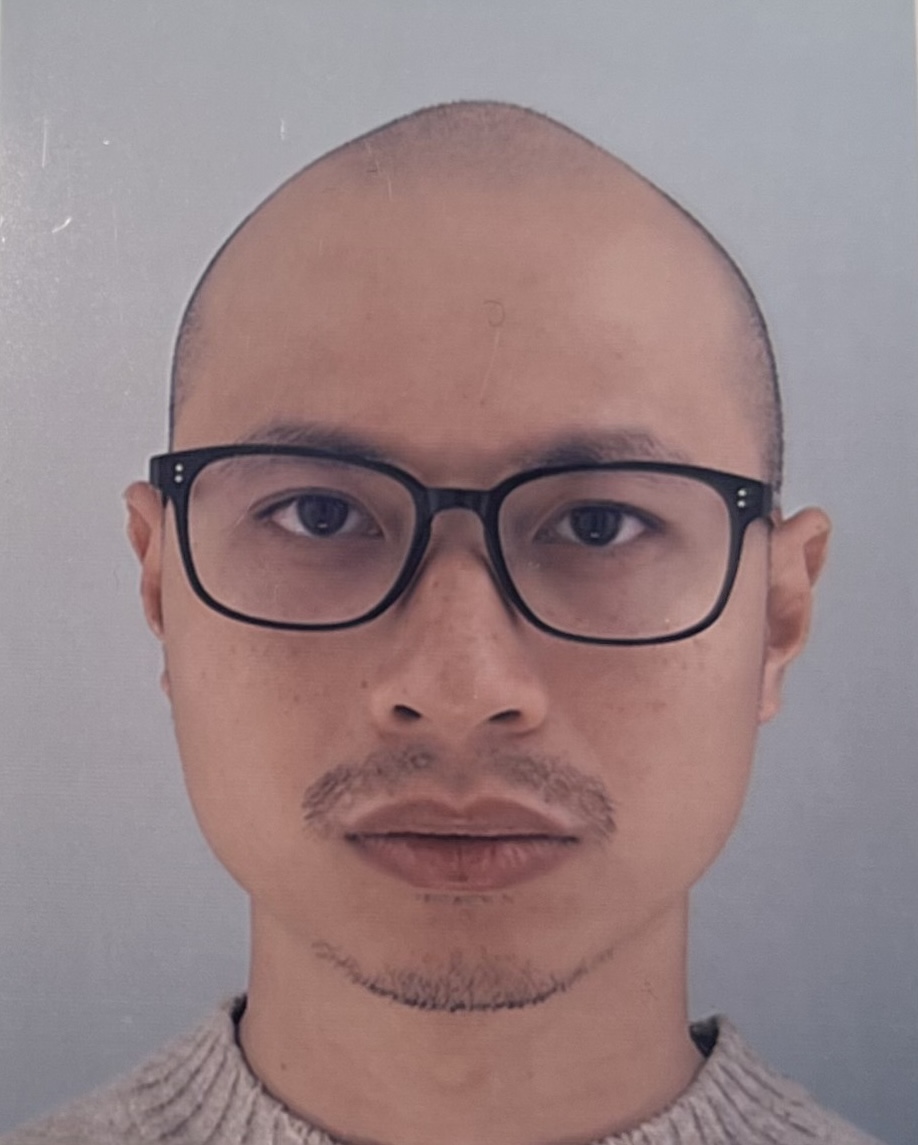}}]{Nam Phuong Tran}
		received the B.S. degree from Hanoi University of Science and Technology, Vietnam, in 2019, and the Ph.D. degree in computer science from the University of Warwick, U.K., in 2025. He is currently a Postdoctoral Researcher with Inria, France. His research interests include decision-making, theoretical artificial intelligence, and its applications for robotics.
	\end{IEEEbiography}

	\begin{IEEEbiography}[{\includegraphics[width=1in,height=1.25in,clip,keepaspectratio]{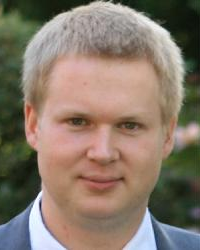}}]{Matthew D. Higgins} (Senior Member, IEEE) is a Professor at the University of Warwick, where he leads WMG's Communications and Navigation Research Group. His research interests span 5G and beyond, core networking, IEEE 802.3xx, GNSS, and timing, with applications to both the automotive and manufacturing domains. Motivated by the need to ensure the ongoing resilience of these domains, he leads many high-value collaborative projects funded through EPSRC, Innovate UK, and HVMC, as well as multiple projects funded directly by industry.
	\end{IEEEbiography}
\end{document}